\documentclass[ reprint,  amsmath,amssymb,  aps, ]{revtex4-1}%
\usepackage{graphicx}
\usepackage{dcolumn}
\usepackage{bm}
\usepackage{amsmath}
\usepackage{amsfonts}
\usepackage{amssymb}
\usepackage{color}

\providecommand{\U}[1]{\protect\rule{.1in}{.1in}}

\begin{document}

\title{
Field-induced valence fluctuations in YbB$_{12}$
}

\author{
R. Kurihara$^1$, 
A. Miyake$^1$,
M. Tokunaga$^1$,
A. Ikeda$^1$,
Y. H. Matsuda$^1$,
A. Miyata$^2$,
D. I. Gorbunov$^2$,
T. Nomura$^{1,2}$,
S. Zherlitsyn$^2$,
J. Wosnitza$^2$,
and
F. Iga$^3$
}

\affiliation{
$^1$Institute for Solid State Physics, The University of Tokyo, Kashiwa, Chiba 277-8581, Japan \\
$^2$Hochfeld-Magnetlabor Dresden (HLD-EMFL) and W\"urzburg-Dresden Cluster of Excellence ct.qmat, Helmholtz-Zentrum Dresden-Rossendorf, 01328 Dresden, Germany \\
$^3$College of Science, Ibaraki University, Mito 310-8512, Japan
}

\begin{abstract}
We performed high-magnetic-field ultrasonic experiments on YbB$_{12}$ up to 59 T to investigate the valence fluctuations in Yb ions.
In zero field, the longitudinal elastic constant $C_{11}$, the transverse elastic constants $C_{44}$ and $\left( C_{11} - C_{12} \right)/2$, and the bulk modulus $C_\mathrm{B}$ show a hardening with a change of curvature at around 35 K indicating a small contribution of valence fluctuations to the elastic constants.
When high magnetic fields are applied at low temperatures, $C_\mathrm{B}$ exhibits a softening above a field-induced insulator-metal transition signaling field-induced valence fluctuations.
Furthermore, at elevated temperatures, the field-induced softening of $C_\mathrm{B}$ takes place at even lower fields and $C_\mathrm{B}$ decreases continuously with field.
Our analysis using the multipole susceptibility based on a two-band model reveals that the softening of $C_\mathrm{B}$ originates from the enhancement of multipole-strain interaction in addition to the decrease of the insulator energy gap.
This analysis indicates that field-induced valence fluctuations of Yb cause the instability of the bulk modulus $C_\mathrm{B}$.

\end{abstract}

\maketitle
%%%%%%%%%%%%%%%%Affil%%%%%%%%%%%%%%%%%%%%%%%%%%%%%

%%%%%%%%%%%%%%%%ABST%%%%%%%%%%%%%%%%%%%%%%%%%%%%

%%%%%%%%%%%%%%%%%%%%%    Introduction     %%%%%%%%%%%%%%%%%%%%%%%%%%%%%%%%%%
\section{
\label{sect_intro}
Introduction
}
%%%high field high magnetic field
Since the electronic and magnetic properties of materials are mainly determined by valence electrons, a precise knowledge about the valence state is important in material science.
Especially for $4f$-electron systems, the valence determines the total angular momentum $J$, the localized (or delocalized) $4f$-electron character, and corresponding wave functions.
A non-integer valence state appears in some rare-earth compounds with Ce, Sm, Eu, and Yb ions. In such materials, valence fluctuations due to hybridization between conduction electrons and $4f$ electrons play a key role in their physical properties.
YbB$_{12}$ is one of the valence fluctuating materials with such a $c$-$f$ hybridization, a high-Kondo temperature, and insulating character
\cite{Kasaya_JMMM31, Susaki_PRL77}.

YbB$_{12}$ has the UB$_{12}$-type crystal structure belonging to the $Fm\overline{3}m$ ($O_h^5$) space group 
\cite{Kasaya_JMMM31}.
The $\Gamma_8$ ground state of the $4f$ electrons based on Yb$^{3+}$ configuration in the crystal electric field (CEF) has been proposed
\cite{Nemkovski_PRL99, Kanai_JPSJ84}.
The almost degenerated $\Gamma_7$ and the $\Gamma_6$ states at 270 K (23 meV) were considered as excited states 
\cite{Nemkovski_PRL99, Kanai_JPSJ84}.
These CEF states based on the $J = 7/2$ can be consistent with the hyperfine coupling constant for free Yb$^{3+}$ ions determined by NMR measurements
\cite{Ikushima_PhysB281}.
In contrast, a nonmagnetic ground state has been suggested from the temperature-independent magnetic susceptibility at low temperatures \cite{Kasaya_JMMM47, Iga_JMMM177}.
Indication for a strongly hybridized electronic state was found using bulk-sensitive x-ray photoelectron spectroscopy showing a slight deviation from the valence Yb$^{3+}$
\cite{Yamaguchi_PRB79}.
The hybridization between $5d$ conduction electrons and $4f$ localized electrons has been proposed as a candidate mechanism for an observed band-gap opening
\cite{Saso_JPSJ72, Ohashi_PRB70}.
The contribution of the B-$2p$ electrons to the $c$-$f$ hybridization is also discussed as a result of $dd\sigma$ hopping through B$_{12}$ clusters.

In addition to the CEF scheme, several characteristic energies related to the insulating character have been studied in YbB$_{12}$.
Both in a polycrystal and single crystal, resistivity measurements show evidence for two activation energies of $\sim$ 30 and 65 K 
\cite{Kasaya_JMMM47, Iga_JMMM177}.
A density of states with two-double peaks was proposed as a mechanism of two activation energies
\cite{Sugiyama_JPSJ57}.
NMR and specific-heat data have been described by a simple two-band model, each band having a bandwidth of 55 K, and with an energy gap of 140 K at the Fermi energy
\cite{Kasaya_JMMM47, Iga_JMMM76}.
High-resolution photoemission spectroscopy suggested a hybridization gap of 170 K (15 meV) below 150 K and strongly hybridized character below 60 K
\cite{Takeda_PRB73}.

In YbB$_{12}$, various high-magnetic-field studies were performed to elucidate the mechanism of the formation of the energy gap.
High-field magnetoresistance measurements indicated that the energy gap of 30 K closes around 45 T while the other gap remains up to higher fields
\cite{Sugiyama_JPSJ57}.
Magnetization measurements revealed metamagnetic behavior at insulator-metal (IM) transitions at $B_\mathrm{IM} = 47$ T for $B\|[001]$ and 54 T for $B\|[110]$ and $B\|[111]$
\cite{Iga_JPhysConfSer200}.
Another magnetization anomaly indicating the saturation of magnetization appears at 102 T
\cite{Terashima_JPSJ86}.
The energy shift of the $4f$ band due to the Zeeman effect was proposed to explain the closing of the band gap of 170 K.
Synchrotron x-ray absorption spectra showed the field independence of the $L_3$ edge indicating no considerable change of the Yb valence in the field-induced metal phase
\cite{Matsuda_JPhys51}.
Specific-heat measurements revealed a discontinuous enhanced of Sommerfeld coefficient, $\gamma \sim 60$ mJ/mol$\cdot$K$^2$, and a corresponding Kondo temperature of $220$ - $250$ K above the IM transition, suggesting that the high-field phase is a valence-fluctuating Kondo metal
\cite{Terashima_PRL120}.
These high-field experiments indicate a contribution of the $c$-$f$ hybridization to the opening of the energy gap in YbB$_{12}$.
Magnetic quantum oscillations in the insulating phase have also been focused to understand the insulating character of YbB$_{12}$
\cite{Xiang_Science362}.

\begin{figure}[t]
\begin{center}
\includegraphics[clip, width=0.5\textwidth, bb=0 0 530 260]{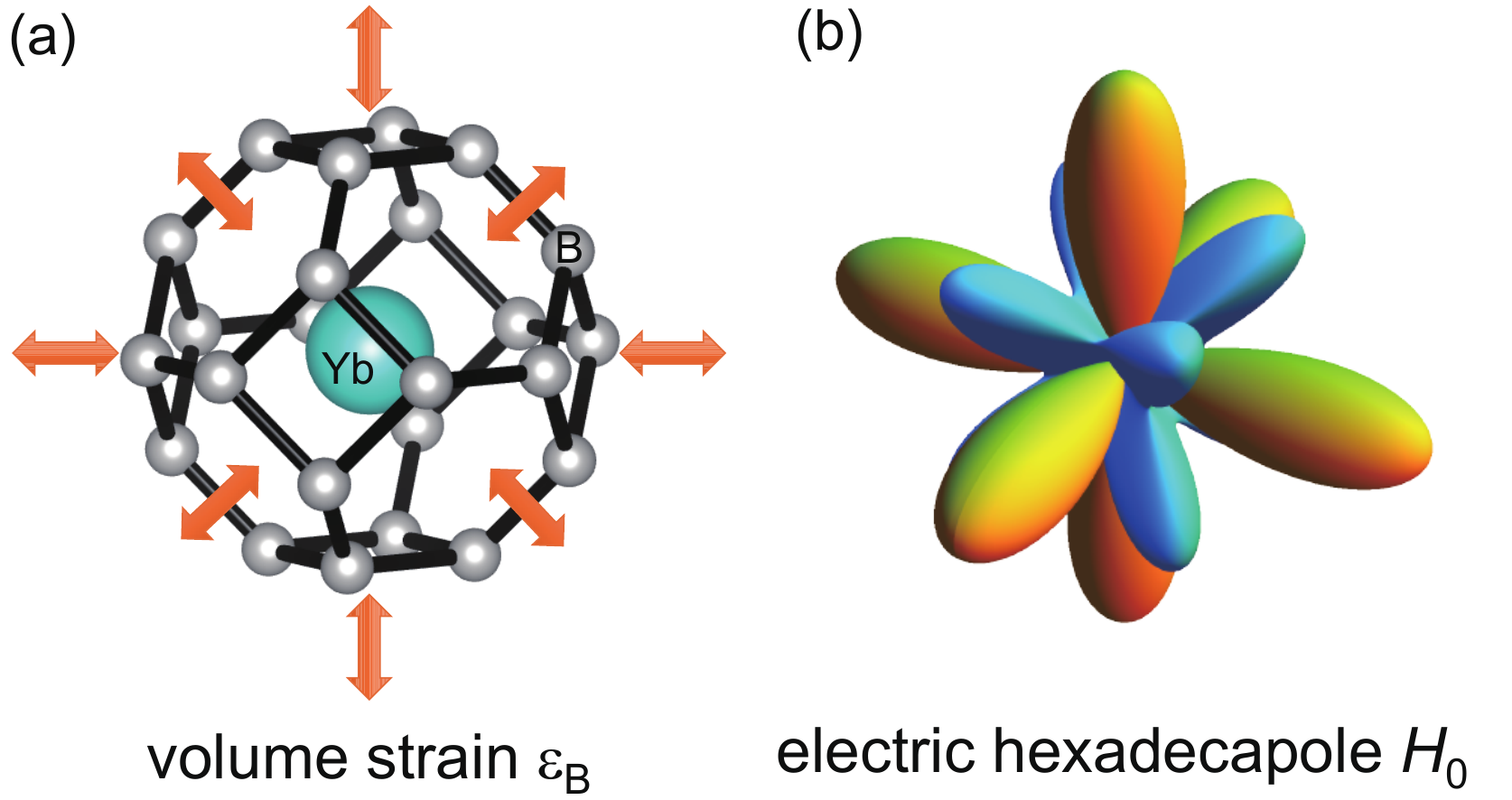}
\end{center}
\caption{
Schematic view of the volume strain and the hexadecapole in YbB$_{12}$.
(a) Crystal lattice around the Yb ion \cite{Kanai_JPSJ84} and volume strain $\varepsilon_\mathrm{B}$ with the irrep $\Gamma_1$ of $O_h$.
Orange arrows indicate the isotropic deformation of the lattice.
(b) Hexadecapole $H_0$ with $\Gamma_1$ symmetry obtained as a result of an isotropic change of the Yb ionic radius.
}
\label{Fig1}
\end{figure}

To further investigate the valence fluctuations caused by the $c$-$f$ hybridization in YbB$_{12}$, we focused on ultrasonic measurement.
Since a valence change causes an isotropic change of the ionic radii, an isotropic volume change of the crystal lattice is induced and the CEF Hamiltonian, $H_\mathrm{CEF}$, of Eq. (\ref{H_CEF}) (see Appendix \ref{Appendix_B}) is changed to $H_\mathrm{CEF} + ( \partial H_\mathrm{CEF} / \partial \varepsilon_\mathrm{B}) \varepsilon_\mathrm{B}$.
Here, $\varepsilon_\mathrm{B}$ is the volume strain with the irreducible representation (irrep) $\Gamma_1$ of the $O_h$ symmetry.
This additional term to the CEF is described as a coupling between $\varepsilon_\mathrm{B}$ and a hexadecapole $H_0$ with $\Gamma_1$ in YbB$_{12}$.
The schematic view of $\varepsilon_\mathrm{B}$ and $H_0$ in YbB$_{12}$ are shown in Figs. \ref{Fig1}(a) and 1(b), respectively.
Based on simple Landau theory for elasticity, the total free energy consists of a lattice and an electronic part is given by \cite{Luthi Phys. Ac.}
\begin{align}
\label{Landau}
F = \frac{1}{2}C_\mathrm{B}^0 \varepsilon_\mathrm{B}^2 + \frac{1}{2} \alpha H_0^2 - g_\mathrm{B}H_0 \varepsilon_\mathrm{B}
.
\end{align}
Here, $g_\mathrm{B}$ is the coupling constant between the strain and $H_0$, $C_\mathrm{B}^0$ is the bulk modulus without multipole contribution, and $\alpha$ is a coefficient.
The 1st and 2nd terms on the right-hand side of Eq. (\ref{Landau}) correspond to the energy loss due to the deformation of the lattice and the increase of the hexadecapole moment, respectively.
The 3rd term corresponds to the energy gain of the electronic state due to the hexadecapole-volume strain interaction.
The response of the hexadecapole appears as a result of the decrease in the bulk modulus as $C_\mathrm{B}^0 - g_\mathrm{B}^2/\alpha$.

%%%%%%%%%%%%%%%%%%%%%%%%Fig1%%%%%%%%%
%%%%%%%%%%%%%%%%%%%%%%%%%%%%%%

As shown in previous reports \cite{Tamaki_JPhysC18, Nemoto_PRB61, Goto_PRB59},
ultrasonic measurements are a powerful tool to detect valence fluctuations. In particular, in the Kondo insulator SmB$_6$, the decrease in the bulk modulus $C_\mathrm{B}$ with decreasing temperatures, namely the elastic softening of $C_\mathrm{B}$, has been revealed as a result of valence fluctuations between Sm$^{2+}$ and Sm$^{3+}$
\cite{Nakamura_JPSJ60_SmB6}.
The relation between the energy gap of $c$-$f$ hybridized bands and the elastic softening is also discussed in terms of the interaction between $4f$ electrons and the bulk strain $\varepsilon_\mathrm{B}$ with full symmetry $\Gamma_1$.
Several theoretical studies have proposed such a contribution of the $c$-$f$ hybridization to the elasticity
\cite{Luthi_JMMM63, Thalmeier_JPhysC20, Keller_PRB41, Rout_PhysicaB367}.
Therefore, we measured relevant elastic constants in zero and high fields searching for an elastic softening related to the valence fluctuations in YbB$_{12}$.

This paper is organized as follows. In Sec. \ref{sect_exp}, experimental details of sample preparation and ultrasonic measurements in pulsed magnetic fields are explained.
In Sec. \ref{Sect_3}, we present the results of our ultrasonic experiments of YbB$_{12}$.
In zero field, an increase in the elastic constants with decreasing temperatures, namely elastic hardening, accompanying curvature changes reveals some contribution of valence fluctuations to the elasticity.
In contrast to zero field, a field-induced softening of the bulk modulus $C_\mathrm{B}$ appears, which indicates field-induced valence fluctuations due to $c$-$f$ hybridization.
In Sec. \ref{discussion}, we analyzed the measured elastic constants using a multipole-susceptibility model. The field-induced valence fluctuations can be described in terms of the hexadecapole-volume strain coupling.
Our analysis also confirms the decrease of the energy gap in high fields.
We summarize our results in Sec. \ref{conclusion}.

%%%%%%%%%%%%%%%%%%%%%    Experiment     %%%%%%%%%%%%%%%%%%%%%%%%%%%%%%%%%%
\section{
\label{sect_exp}
Experiment
}

Single crystals of YbB$_{12}$ were grown using the floating-zone method
\cite{Iga_JMMM177}.
Laue x-ray backscattering was used to align, cut, and polish samples with (110), (1$\bar{1}$0), ($\bar{1}$10), ($\bar{1}$$\bar{1}$0), (001), and (00$\bar{1}$) faces and the size of $1.033$ mm $\times 1.030$ mm $\times 3.763$ mm.
An ultrasound pulse-echo method with a numerical vector-type phase-detection technique was used to measure the ultrasound velocity $v$
\cite{Fujita_JPSJ80}.
The elastic constant $C = \rho v^2$ was determined from $v$ and the calculated mass density $\rho = 4.828$ g/cm$^3$ using the lattice constant $a = 7.469 \mathrm{\AA}$
\cite{Kasaya_JMMM31}.
Piezoelectric transducers using LiNbO$_3$ plates with a 36$^\circ$ Y-cut and 41$^\circ$ X-cut (YAMAJU CERAMICS CO.) were employed to generate longitudinal ultrasonic waves with the fundamental frequency of approximately $f$ = 30 MHz and transverse waves with 18 MHz, respectively. 
As indicated in Fig. \ref{Fig2}, higher-harmonic frequencies were used to obtain high-resolution data.
A room temperature vulcanizing rubber (Shin-Etsu Silicone KE-42T) was used to glue the LiNbO$_3$ on the sample.
The direction of ultrasonic propagation, $\boldsymbol{q}$, and the direction of polarization, $\boldsymbol{\xi}$, for the elastic constant $C_{ij}$ are indicated in Fig. \ref{Fig2}.
Two nondestructive pulsed magnets were used:
one with a pulse duration of 36 ms installed at the Institute for Solid State Physics, the University of Tokyo using a $^4$He cryostat, and another magnet with a pulse duration of 150 ms at the HLD-EMFL in Dresden using a $^3$He cryostat.

%%%%%%%%%%%%%%%%%%%%%   Results    %%%%%%%%%%%%%%%%%%%%%%%%%%%%%%%%%%
\section{
\label{Sect_3}
Results
}

\subsection{
\label{TempDep}
Temperature dependence of elastic constants
}

\begin{figure}[htbp]
\begin{center}
\includegraphics[clip, width=0.5\textwidth, bb=0 0 400 680]{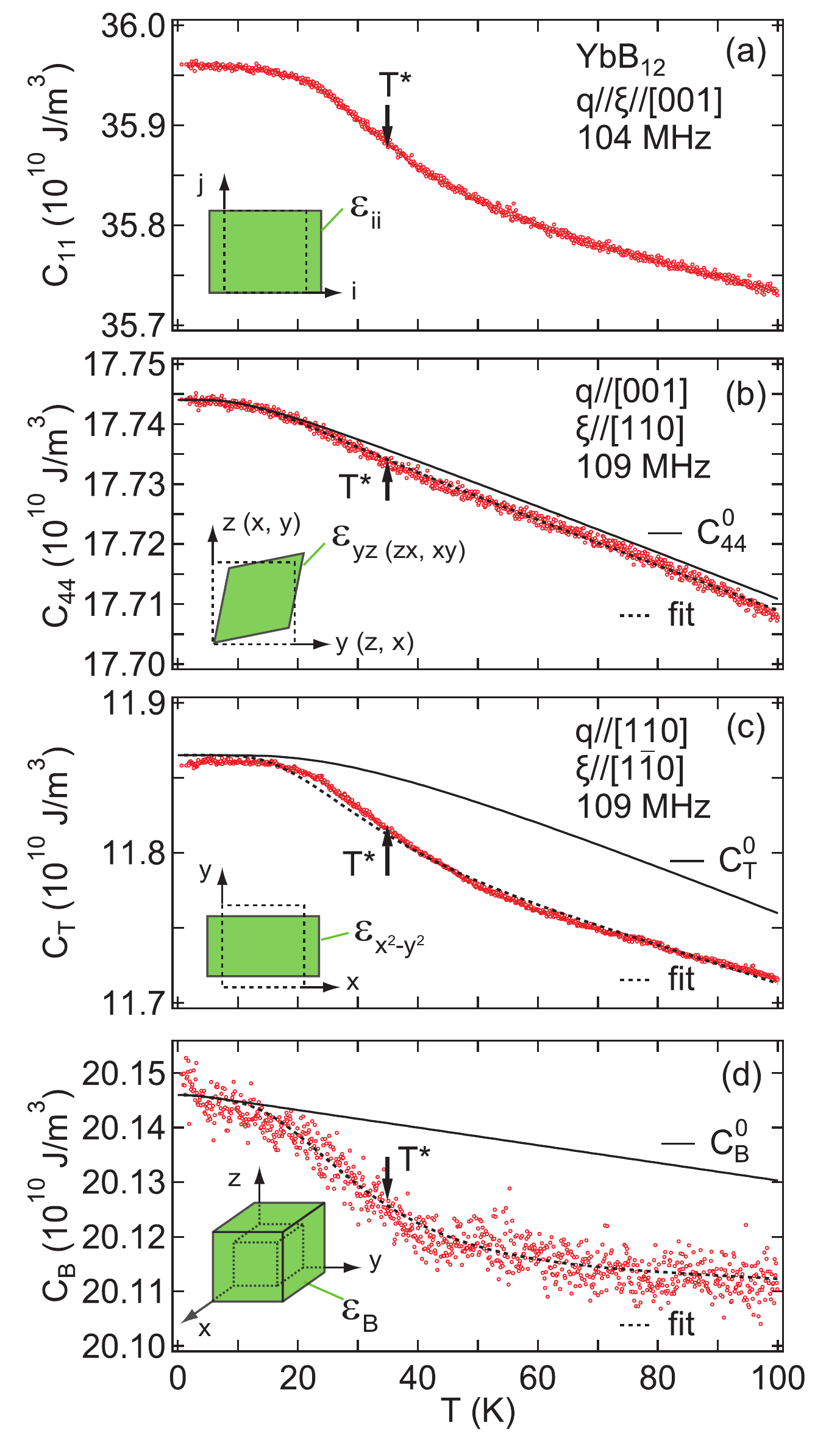}
\end{center}
\caption{
Temperature dependence of (a) the longitudinal elastic constant $C_{11}$, the transverse elastic constants (b) $C_{44}$ and (c) $C_\mathrm{T} = \left( C_{11} - C_{12} \right)/2$, and (d) the bulk modulus $C_\mathrm{B}$ calculated from  $C_{11}$ and $C_\mathrm{T}$.
The dotted lines indicate the fit of $C_{44}$, $C_\mathrm{T}$, and $C_\mathrm{B}$ in the framework of the phenomenological two-band model discussed in Sec. \ref{discussion}.
The solid lines indicate the temperature dependence of the elastic constants without multipole contribution.
The strains $\varepsilon_\mathrm{ij}$ for $C_{11}$, $C_{44}$, $C_\mathrm{T}$, and $C_\mathrm{B}$ are schematically shown in the inset in each panel.
The vertical arrows in each panel indicate the characteristic temperature $T^\star$ discussed in Sec. \ref{discussion}. 
}
\label{Fig2}
\end{figure}

To gain more information on the Yb valence in YbB$_{12}$, we investigated the three elastic constants $C_{11}$, $C_{44}$, and $C_\mathrm{T} = \left( C_{11} - C_{12} \right)/2$.
Their relations to the symmetry strain and electric multipole are summarized in Table \ref{table1}
\cite{Inui_group, Luthi Phys. Ac.}.
Figure \ref{Fig2} shows the temperature dependence of the elastic constants in zero field.
We observed the elastic hardening of $C_{11}$, $C_{44}$, and $C_\mathrm{T}$ with lowering temperatures.
We also observed the elastic hardening of $C_{11}$ from 300 K (see Appendix \ref{Appendix_A}).
All elastic constants exhibit an additional hardening and a characteristic curvature change in the vicinity of $T^\star = 35$ K.
As shown by the solid curves in Fig. \ref{Fig2}, the elastic constants would exhibit a monotonic increase with decreasing temperature 
\cite{Vershni_PRB2}
if we do not consider multipole contributions, described in the following Sec. \ref{discussion}
\cite{Luthi Phys. Ac.}.
Therefore, the additional features in the elastic constants of YbB$_{12}$ indicate the multipole contribution to elasticity.

%%%%%%%%%%%%%%%%%%%%%%%%Fig2%%%%%%%%%
%%%%%%%%%%%%%%%%%%%%%%%%%%%%%%%

To describe the origin of the anomaly in each elastic constant of YbB$_{12}$, we focus on the multipole effect of the CEF wave functions of localized $4f$ electrons taken into account the presence of $\Gamma_8$, $\Gamma_7$, and $\Gamma_6$ states 
\cite{Nemkovski_PRL99, Kanai_JPSJ84}.
Since the direct product of the $\Gamma_8$ quartet is reduced as $\Gamma_8 \otimes \Gamma_8 
= \Gamma_1 \oplus \Gamma_2 \oplus \Gamma_3 \oplus 2\Gamma_4 \oplus 2\Gamma_5$
\cite{Inui_group, Kuramoto_JPSJ78},
we deduce that the $\Gamma_8$ ground-state wave functions carry the electric quadrupoles 
$O_{u}$ and $O_{v}$ with irrep $\Gamma_3$
and 
$O_{yz}$, $O_{zx}$, and $O_{xy}$ with irrep $\Gamma_5$ as summarized in Table \ref{table1}.
In addition, the $\Gamma_8$ quartet also provides the electric hexadecapole
$H_0 = O_4^0 + 5O_4^4$ with irrep $\Gamma_1$.
Because the magnetic multipole degrees of freedom do not couple with the strain, we ignore magnetic dipoles with irrep $\Gamma_4$ and magnetic octupoles with irreps $\Gamma_2$, $\Gamma_4$, and $\Gamma_5$. 
This group-theoretical consideration indicates that the elastic softening of $\left( C_{11}-C_{12} \right)/2$ with irrep $\Gamma_3$ and $C_{44}$ with irrep $\Gamma_5$ is due to a multipole-strain interaction described as
%%%%%%%%H_MS%%%%%%%%%%%%%
\begin{align}
\label{H_MS}
H_\mathrm{MS}
= -g_{\Gamma_\gamma} O_{\Gamma_\gamma} \varepsilon_{\Gamma_\gamma}
.
\end{align}
Here, $g_{\Gamma_\gamma}$ is a coupling constant and $\Gamma_\gamma$ denotes the irrep.
We show how to calculate the multipole susceptibility based on the CEF wave functions in Appendix \ref{Appendix_B}.
Because the calculated multipole susceptibility for $\Gamma_3$- and $\Gamma_5$-type quadrupoles shows a divergent increase for decreasing temperatures, a divergent elastic softening is theoretically expected in $C_{44}$ and $C_\mathrm{T}$.
However, our experimental results show no softening in all measured elastic constants.
Therefore, the CEF approach based on a localized $4f$ character does not apply to the elasticity of YbB$_{12}$ in zero field.

The other possible scenario describing the additional contribution around $T^\star$ is a result of the charge freezing of Yb without long-range ordering as previously discussed in the samarium compounds Sm$_3$Se$_4$ and Sm$_3$Te$_4$
\cite{Tamaki_JPhysC18, Nemoto_PRB61}.
Since the charge freezing would be characterized by a frequency-dependent ultrasound response, we measured the elastic constants and ultrasonic attenuation coefficients for several frequencies.
However, we did not observe any frequency dependence neither in the elastic constants nor in the ultrasonic attenuation coefficients between 30 and 160 MHz.

%%%%%%%%%%%%%%%%Table1%%%%%%%%%%%%%%%%%
\begin{table*}[t]
\caption{
Symmetry strains, electric multipoles, and elastic constants corresponding to the irreducible representations (irreps) of the space group $O_h$.
}
\begin{ruledtabular}
\label{table1}
\begin{tabular}{cccccc}
\textrm{Irrep}
& \textrm{Symmetry strain}
	& \textrm{Electric multipole}
		& \textrm{Elastic constant}
\\
\hline
$\Gamma_1$
	& $\varepsilon_\mathrm{B} = \varepsilon_{xx} + \varepsilon_{yy} + \varepsilon_{zz}$
		& $O_4^0 + 5 O_4^4 (=H_0)$
			& $ C_\mathrm{B} = \left( C_{11} + 2C_{12} \right) /3 $
\\
$\Gamma_3$ 
	& $\varepsilon_u = (2\varepsilon_{zz} - \varepsilon_{xx} - \varepsilon_{yy})/\sqrt{3}$
		& $O_{u(=3z^2-r^2)}$
			& $C_\mathrm{T} = \left( C_{11} - C_{12} \right)/2$
\\

	& $\varepsilon_v = \varepsilon_{xx} - \varepsilon_{yy}$
		& $O_{v(=x^2-y^2)}$
			&
\\
$\Gamma_5$
	& $\varepsilon_{yz}$
		& $O_{yz}$
			& $C_{44}$
\\

	& $\varepsilon_{zx}$
		& $O_{zx}$
			& 
\\

	& $\varepsilon_{xy}$
		& $O_{xy}$
			&
\end{tabular}
\end{ruledtabular}
\end{table*}
%%%%%%%%%%%%%%%%%%%%%%%%%%%%%%%%%%%

%%%%%%%%%%%%%%%%%%%%%%%%Fig3%%%%%%%%%
\begin{figure*}[htbp]
\begin{center}
\includegraphics[clip, width=1\textwidth, bb=0 0 650 390]{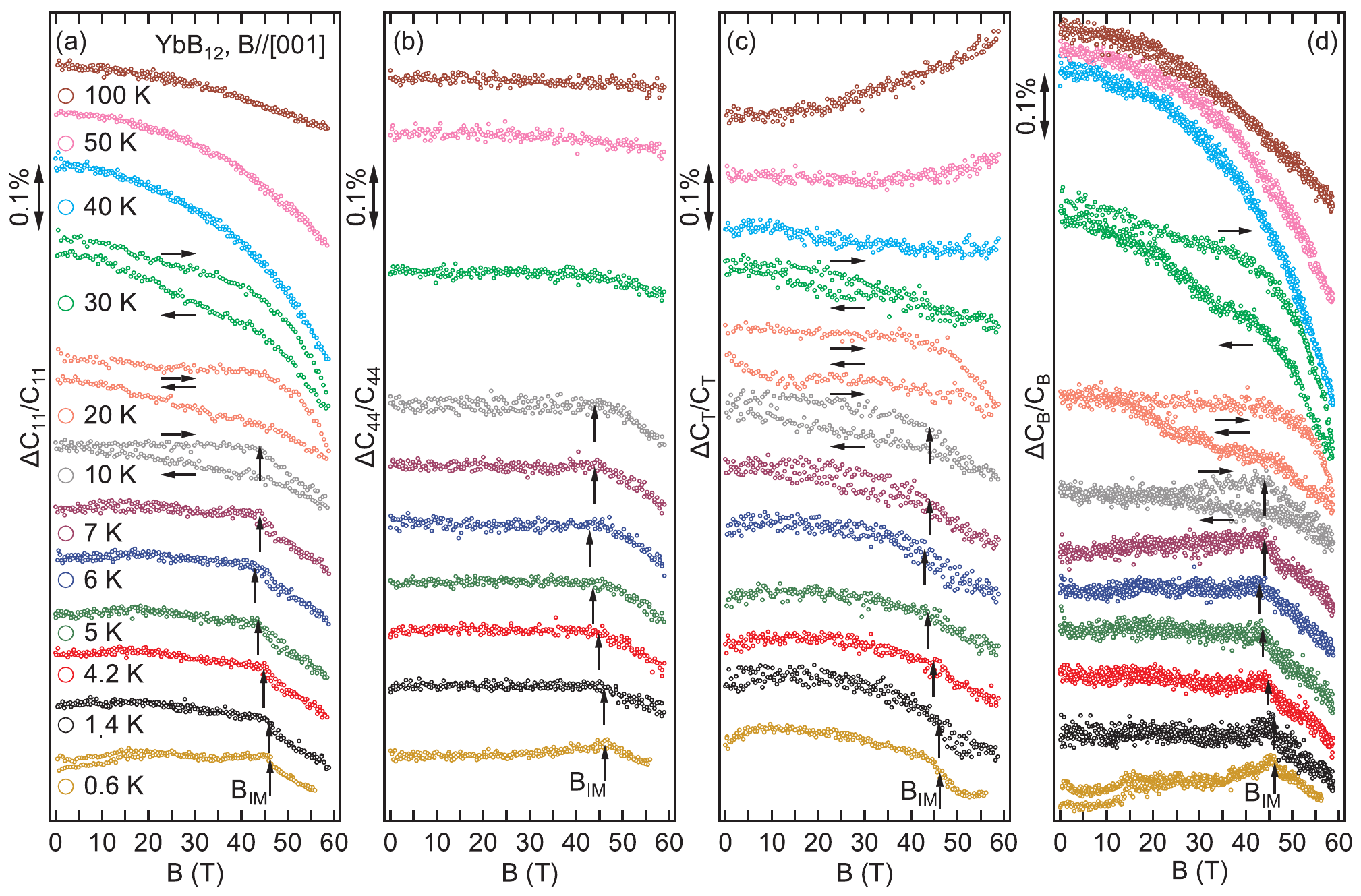}
\end{center}
\caption{
Magnetic-field dependence of the relative variation of the elastic constants
$\Delta C_{ij}/C_{ij} = \left[ C_{ij}(B) - C_{ij}(B = 0) \right] /C_{ij}(B = 0)$
at several temperatures for $B\|[001]$.
Field dependence of (a) the longitudinal elastic constant $C_{11}$, the transverse elastic constants (b) $C_{44}$ and (c) $C_\mathrm{T}$, and (d) the bulk modulus $C_\mathrm{B}$.
The data sets are shifted consecutively along the $\Delta C/C$ axes for clarity.
The vertical arrows indicate the insulator-metal transition field $B_\mathrm{IM}$.
The horizontal arrows show the field-sweep directions.
}
\label{Fig3}
\end{figure*}
%%%%%%%%%%%%%%%%%%%%%%%%%%%%%%%%%%%%%%%%%%%%%%%%%%%

Therefore, we focus on the contribution of valence fluctuations to the elasticity caused by the $c$-$f$ hybridization.
Figure \ref{Fig2}(d) shows the temperature dependence of the bulk modulus $C_\mathrm{B} = ( C_{11} + 2C_{12} )/3 = C_{11} -4C_\mathrm{T}/3$ with the irrep $\Gamma_1$ calculated from the experimental results of $C_{11}$ and $C_\mathrm{T}$.
$C_\mathrm{B}$ exhibits as well a hardening with an additional contribution in the vicinity of 35 K.
This result for YbB$_{12}$ is in contrast to the significant softening of $C_\mathrm{B}$ due to Sm valence fluctuations observed in SmB$_6$.
In Sec. \ref{discussion}, we will discuss the origin of the additional contribution in terms of the multipole susceptibility based on a two-band model to confirm the contribution of valence fluctuations to the elastic constants in zero field.

%%%%%%%%%%%%%%%%%%%%%%%%%%%%%%%%%%%%%%%%%%%%%%%%%%

\subsection{
\label{Bdep}
Magnetic-field dependence of elastic constants
}

To investigate the valence properties of YbB$_{12}$ in magnetic fields, we measured the elastic constants $C_{11}$, $C_{44}$, and $C_\mathrm{T}$ up to 59 T for $B\|[001]$.
Figure \ref{Fig3} shows the magnetic-field dependence of the relative variation of the elastic constants $\Delta C_{ij}/C_{ij}$ at several temperatures.
We observed a field-induced IM transition and elastic softening for each elastic constant in the Kondo-metal phase.
Below 10 K, this softening appears rather abruptly above the insulator-metal transition field $B_\mathrm{IM}$.
$B_\mathrm{IM}$ are comparable with results of a previous magnetocaloric-effect study
\cite{Terashima_PRL120}.
Since $C_\mathrm{B}$ contains $C_{11}$ and $C_\mathrm{T}$, our experimental results show as well the softening of $C_\mathrm{B}$ in the Kondo-metal phase.

Above 10 K,  no sharp anomaly corresponding to the Kondo-metal phase transition is visible any more.
However, $C_{11}$ still shows a significant softening in magnetic field contrary to the other elastic constants (Fig. \ref{Fig3}).
In particular, at 40 K, $C_{11}$ exhibits a large softening of $0.30\%$ at 59 T while $C_\mathrm{T}$ shows a softening of only $0.027\%$.
The softening of $C_{11}$ at 100 K is also in contrast to the hardening observed for $C_\mathrm{T}$.

Between 10 and 30 K, a clear hysteresis appears in the pulsed-field data of $C_{11}$ and $C_\mathrm{T}$ (Fig. \ref{Fig3}).
This is approximately the temperature range where the additional contribution to the elastic constants is detected (Fig \ref{Fig2}). 
As shown in a previous magnetocaloric-effect study in adiabatic condition below 7 K
\cite{Terashima_PRL120},
the temperature of the sample is reduced by the application of a magnetic field.
Because of the quasi-adiabatic experimental conditions, the final temperature after the field pulse fields might be higher than assumed which may cause the hysteresis.
Therefore, the hysteresis of elastic constants $C_{11}$ and $C_\mathrm{T}$ can also be attributed to the magnetocaloric effect.

We also looked for the quantum oscillation in YbB$_{12}$
\cite{Xiang_Science362}.
In principle, such quantum oscillations may appear as well in bulk sensitive ultrasound properties.
However, we were not able to resolve any acoustic de Haas-van Alphen effect at least 0.6 K.
This result may imply a weak electron-phonon interaction for the studied acoustic modes.

%%%%%%%%%%%%%%%%%%%%%%%%Fig4%%%%%%%%%
\begin{figure}[t]
\begin{center}
\includegraphics[clip, width=0.5\textwidth, bb=0 0 450 360]{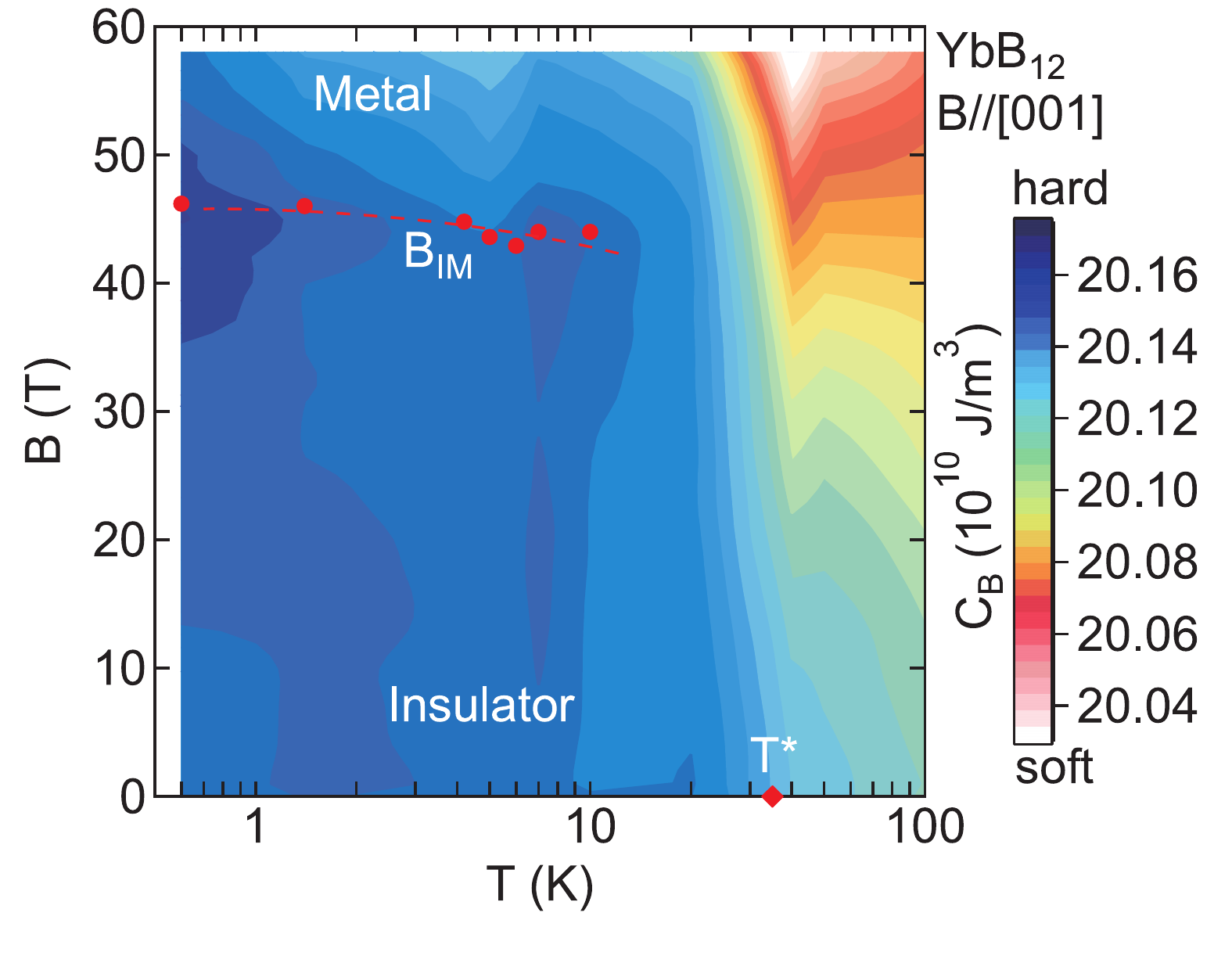}
\end{center}
\caption{
Temperature-field phase diagram of YbB$_{12}$ for $B\|[001]$.
The filled red circles indicate the insulator-metal transition field $B_\mathrm{IM}$.
The filled red diamond indicates the characteristic temperature $T^\star$ (see text for details).
The color code shows the value of the bulk modulus $C_\mathrm{B}$.
}
\label{Fig4}
\end{figure}
%%%%%%%%%%%%%%%%%%%%%%%%%%%%%%%

We calculated $C_\mathrm{B} = C_{11} -4C_\mathrm{T}/3$ from the measured magnetic-field dependence of $C_{11}$ and $C_\mathrm{T}$ (Fig. \ref{Fig3}).
Indeed, $C_\mathrm{B}$ shows a very similar behavior as the individual elastic constants with a clear anomaly at $B_\mathrm{IM}$ below 10 K and hysteresis between 10 and 30 K.
$B_\mathrm{IM}$ determined by our ultrasonic measurements are shown in Fig. \ref{Fig4}. 

$C_\mathrm{B}$ exhibits a small softening below 50 T at 20 K and 45 T at 30 K.
By contrast, above 40 K, $C_\mathrm{B}$ shows monotonic softening with increasing fields.
In particular, the largest softening of 0.52\% is observed in $C_\mathrm{B}$ at 40 K.
The field-induced elastic softening of $C_\mathrm{B}$ is summarized in the contour plot in Fig. \ref{Fig4}. 

%%%%%%%%%%%%%%%%%%%%%%%%Fig5%%%%%%%%%
\begin{figure}[t]
\begin{center}
\includegraphics[clip, width=0.5\textwidth, bb=0 0 440 340]{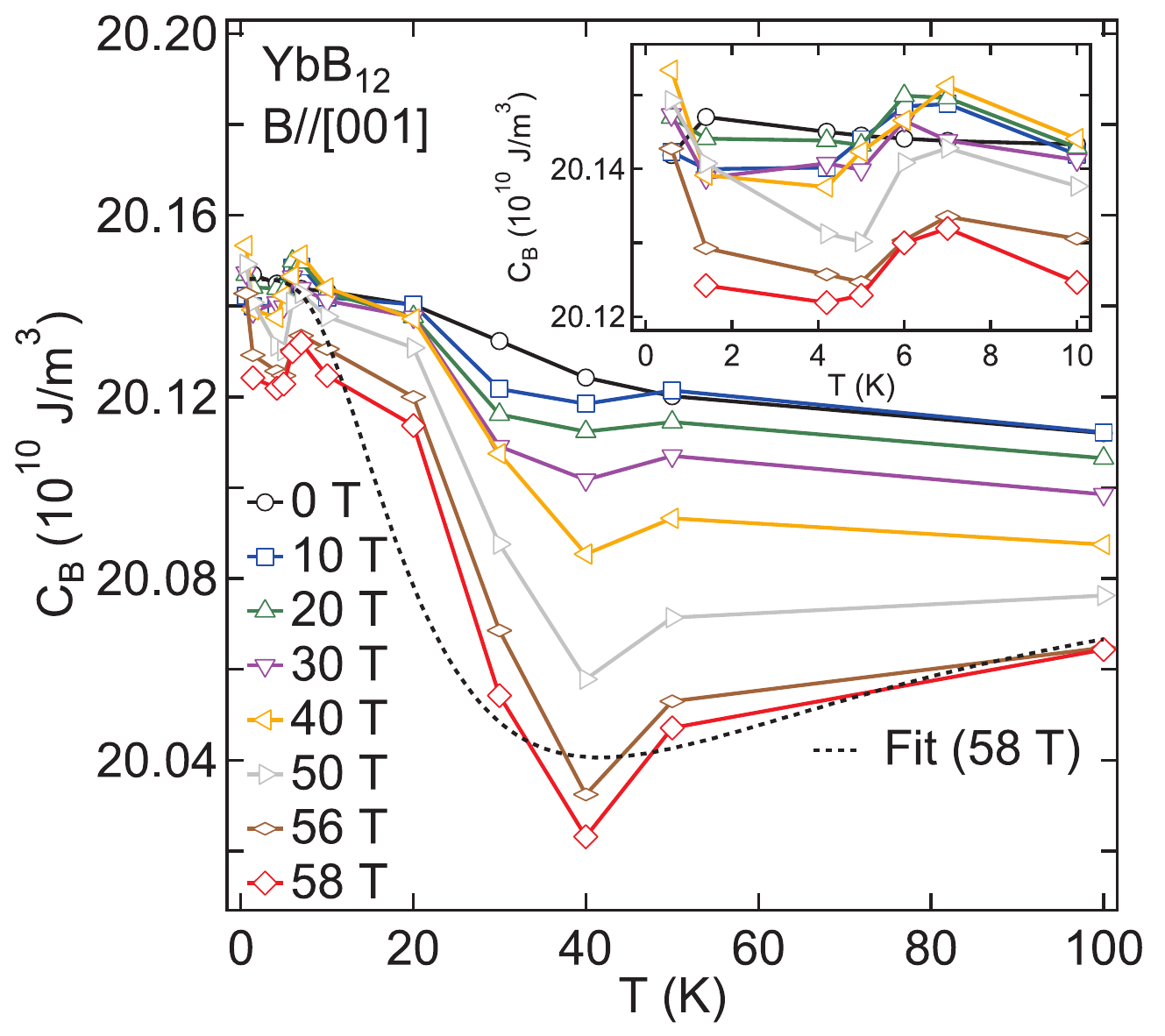}
\end{center}
\caption{
Temperature dependence of the bulk modulus $C_\mathrm{B}$ of YbB$_{12}$ at various magnetic fields for $B\|[001]$ extracted from the up-sweep data shown in Fig. \ref{Fig3}, except for the zero-field data.
The dotted line indicates the fit describing $C_\mathrm{B}$ in the framework of the phenomenological two-band model. The inset shows $C_\mathrm{B}$ below 10 K.
}
\label{Fig5}
\end{figure}
%%%%%%%%%%%%%%%%%%%%%%%%%%%%%%%

For further understanding of the field-induced elastic softening in YbB$_{12}$, we plotted the temperature dependence of $C_\mathrm{B}$ for various magnetic fields (Fig. \ref{Fig5}).
As shown in the inset of Fig. \ref{Fig5}, $C_\mathrm{B}$ exhibits a softening of about $0.05 \%$ below 7 K down to $\sim$ 2 K in magnetic fields above 40 T.
In addition, $C_\mathrm{B}$ shows significant softening from 100 K down to 40 K in high fields, which is in contrast to the hardening of $C_\mathrm{B}$ in zero field.
The softening of $C_\mathrm{B}$ is similar to that found for SmB$_6$ caused by $c$-$f$ hybridization-driven valence fluctuations corresponding to hexadecapole-strain interaction
\cite{Nakamura_JPSJ60_SmB6}.

The experimental results of the temperature dependence and the magnetic field dependence of $C_\mathrm{B}$ of YbB$_{12}$ cannot be described by the localized $4f$-electron model (see Appendix \ref{Appendix_C}).
In the following Sec. \ref{discussion}, therefore, we discuss our observations in terms of multipole-strain interaction and the multipole susceptibility for a two-band model.

%%%%%%%%%%%%%%%%%%discussion%%%%%%%%%%%%%%%

\section{
\label{discussion}
Discussion}

We discuss the origin of the elastic anomalies of YbB$_{12}$ in terms of a two-band model assuming a constant density of states (DOS) with respect to energy.
This model has successfully reproduced the elastic softening observed in the Kondo compounds SmB$_6$ and CeNiSn
\cite{Nakamura_JPSJ60_SmB6, Nakamura_JPSJ60_CeNiSn}. 
In YbB$_{12}$, this phenomenological model also gives qualitative explanation for the temperature dependence of $C_{44}$, $C_\mathrm{T}$, and $C_\mathrm{B}$ in zero field and for $C_\mathrm{B}$ in high fields.
Field-induced valence fluctuations are included by the hexadecapole-strain interaction.
By that, the essential parameters for the explanation of our experimental results are identified (Table \ref{table2}).

%%%%%%%%%%%%%%%%%%%%%%%%Fig6%%%%%%%%%
\begin{figure}[t]
\begin{center}
\includegraphics[clip, width=0.5\textwidth, bb=0 0 650 450]{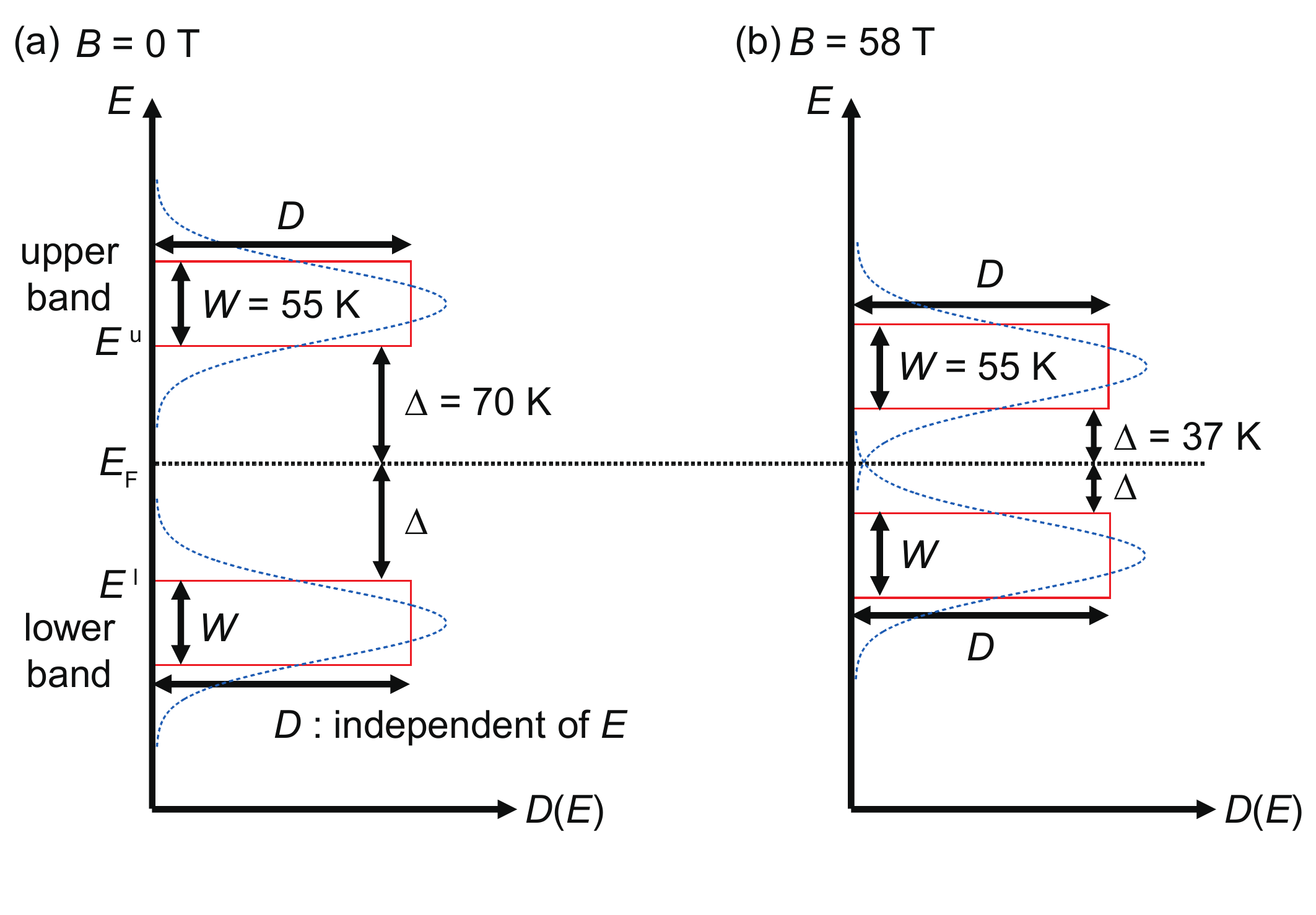}
\end{center}
\caption{
Schematic view of the two-band model assuming a constant DOS over each band (red rectangular).
(a) DOS at zero field with an energy gap $2\Delta = 140$ K and bandwidth $W = 55$ K.
(b) DOS at 58 T. The energy gap $2\Delta = 74$ K and the bandwidth $W = 55$ K is determined using Eq. (\ref{Fitting equation}).
The blue dotted curves indicate the schematic view of a more realistic DOS of YbB$_{12}$.
}
\label{Fig6}
\end{figure}
%%%%%%%%%%%%%%%%%%%%%%%%%%%%%%%%%%%%%%%%%%%%%%%%%%%%%%%%%%%%%

We introduce a two-band model, which is schematically shown in Fig. \ref{Fig6}.
In this model, we deal with the two $c$-$f$ hybridized bands: an upper band above the Fermi energy $E_\mathrm{F}$ with an energy $E_{0, \boldsymbol{k} }^\mathrm{u}$ and a lower band below $E_\mathrm{F}$ with $E_{0, \boldsymbol{k} }^\mathrm{l}$.
The DOS with energy dispersion of each band is simplified to the rectangular form.
The bandwidth $W$, the DOS $D$, and the band gap $2\Delta$ are set as shown in Fig. \ref{Fig6}.
We assume that the multipole-strain interaction for the electrons in the two bands can be written as
\cite{Kurihara_JPSJ86}
%Matrix of O
\begin{align}
\label{HQS_two band model}
H_\mathrm{MS}
= - \sum_{\boldsymbol{k}}
\begin{pmatrix}
c_{\boldsymbol{k}, \mathrm{u} }^\dagger
\cr
c_{\boldsymbol{k}, \mathrm{l} }^\dagger
\end{pmatrix}^T
\begin{pmatrix}
& d_{\boldsymbol{k}, \Gamma_\gamma}^\mathrm{u} & h_{\boldsymbol{k}, \Gamma_\gamma}
\cr
& h_{\boldsymbol{k}, \Gamma_\gamma}^\ast & d_{\boldsymbol{k}, \Gamma_\gamma}^\mathrm{l}
\end{pmatrix}
\begin{pmatrix}
c_{\boldsymbol{k}, \mathrm{u}}
\cr
c_{\boldsymbol{k}, \mathrm{l}}
\end{pmatrix}
\varepsilon_{\Gamma_\gamma} .
\end{align}
For the multipole-strain interaction $H_\mathrm{MS}$ of Eq. (\ref{H_MS}), the diagonal term $d_{\boldsymbol{k}, \Gamma_\gamma}^\mathrm{l(u)}$ for the electrons in band $\mathrm{u(l)}$ indicates a renormalized multipole-strain coupling constant described as
$g_\mathrm{\Gamma_\gamma} \langle \mathrm{u(l)} | O_\mathrm{\Gamma_\gamma} | \mathrm{u(l)} \rangle$.
The off-diagonal term $h_{\boldsymbol{k}, \Gamma_\gamma}$ is written as
$g_\mathrm{\Gamma_\gamma} \langle \mathrm{u(l)} | O_\mathrm{\Gamma_\gamma} | \mathrm{l(u)} \rangle$.
$c_{\boldsymbol{k}, \mathrm{u(l)}}$ and $c_{\boldsymbol{k}, \mathrm{u(l)}}^\dagger$ are annihilation and creation operators of an electron in the band $\mathrm{u(l)}$ with wave vector $\boldsymbol{k}$, respectively.
Considering the Anderson Hamiltonian describing $c$-$f$ hybridization, we deduce that the multipole-strain interaction of Eq. (\ref{HQS_two band model}) originates from electron-phonon interaction consisting of $c$-$f$ and $f$-$f$ terms
\cite{Rout_PhysicaB367}.
The multipole-strain interaction of Eq. (\ref{HQS_two band model}) for the two-band model provides a second-order perturbation for the upper band, the lower one, and the band gap.
The perturbation energies of each band and the perturbation energy gap are described as
\cite{Nakamura_JPSJ60_SmB6, Nakamura_JPSJ60_CeNiSn}
\begin{align}
\label{E_u}
&E_{ \boldsymbol{k} }^\mathrm{u} \left( \varepsilon_{\Gamma_\gamma} \right)
= E_{0, \boldsymbol{k}}^\mathrm{u} -d_{\boldsymbol{k}, \Gamma_\gamma}^\mathrm{u} \varepsilon_{\Gamma_\gamma}
+ \frac{ \left| h_{\boldsymbol{k}, \Gamma_\gamma} \right| ^2 }{2\Delta_{\boldsymbol{k}}} \varepsilon_{\Gamma_\gamma}^2
,
\end{align}
\begin{align}
\label{E_l}
&E_{\boldsymbol{k}}^\mathrm{l} \left( \varepsilon_{\Gamma_\gamma} \right)
= E_{0, \boldsymbol{k}}^\mathrm{l} - d_{\boldsymbol{k}, \Gamma_\gamma}^\mathrm{l} \varepsilon_{\Gamma_\gamma}
- \frac{ \left| h_{\boldsymbol{k}, \Gamma_\gamma} \right| ^2 }{2\Delta_{\boldsymbol{k}}} \varepsilon_{\Gamma_\gamma}^2
,
\end{align}
\begin{align}
\label{Delta_Gap}
&\Delta_{\boldsymbol{k}} \left( \varepsilon_{\Gamma_\gamma} \right)
= \Delta_{ \boldsymbol{k} }
- \frac{1}{2} \left( d_{\boldsymbol{k}, \Gamma_\gamma}^\mathrm{u} - d_{\boldsymbol{k}, \Gamma_\gamma}^\mathrm{l} \right) \varepsilon_{\Gamma_\gamma}
+ \frac{ \left| h_{\boldsymbol{k}, \Gamma_\gamma} \right|^2 }{2 \Delta_{\boldsymbol{k}}} \varepsilon_{\Gamma_\gamma}^2
.
\end{align}
Here,
$2\Delta_{\boldsymbol{k}} = E_{0, \boldsymbol{k}}^\mathrm{u} - E_{0, \boldsymbol{k}}^\mathrm{l}$
is the energy gap between the upper band and the lower one.
The total free energy $F$ is written as
\cite{Luthi_JMMM52}
\begin{align}
\label{Free energy_two band}
F
&= \frac{1}{2}C_{\Gamma_\gamma}^0\varepsilon_{\Gamma_\gamma}^2
+ nE_\mathrm{F} \left( \varepsilon_{\Gamma_\gamma} \right)
\nonumber \\
&- k_\mathrm{B} T \sum_{s ( = \mathrm{u, l}), \boldsymbol{k} }
\ln \left\{ 1 +
\exp \left[
{-\frac{
E_{\boldsymbol{k}}^s\left( \varepsilon_{\Gamma_\gamma} \right) - E_\mathrm{F} \left( \varepsilon_{\Gamma_\gamma} \right)
}{
k_\mathrm{B} T}
}
\right] \right\}
.
\end{align}
Here, $C_{\Gamma_\gamma}^0$ is the elastic constant due to the phonon part with the irrep $\Gamma_\gamma$,
$n$ is the total number of conduction electrons, $E_\mathrm{F} (\varepsilon_{\Gamma_\gamma})$ is the Fermi energy in the deformed system, and $k_\mathrm{B}$ is the Boltzmann constant.
The first term on the right-hand side of Eq. (\ref{Free energy_two band}) corresponds to the lattice part.
The second and third terms correspond to the free energy of the conduction electrons.
The second derivative of the total free energy with respect to the strain $\varepsilon_\Gamma$ provides the elastic constant $C_{\Gamma_\gamma}(T)$ described as
%%%%%%%%%%%%%%%%%%%%%%%%%%%%%%%%elastic band model
\begin{align}
\label{Elastic_Band Model}
C_{\Gamma_\gamma}(T)
&= C_{\Gamma_\gamma}^0
+ \sum_{s, \boldsymbol{k}}
\frac{ \partial^2 E_{\boldsymbol{k}}^s }{ \partial \varepsilon_{\Gamma_\gamma}^2 }
f_{\boldsymbol{k}}^s
\nonumber \\
&-\frac{1}{k_\mathrm{B}T}\sum_{s, \boldsymbol{k}}
\left( \frac{\partial E_{\boldsymbol{k}}^s }{\partial \varepsilon_{\Gamma_\gamma}} \right)^2
f_{\boldsymbol{k}}^s \left( 1-f_{\boldsymbol{k}}^s \right)
\nonumber \\
&+\frac{1}{k_\mathrm{B}T}
\frac{
\sum_{s, \boldsymbol{k}}
\left[ \frac{\partial E_{\boldsymbol{k}}^s }{\partial \varepsilon_{\Gamma_\gamma}}
f_{\boldsymbol{k}}^s \left( 1 - f_{\boldsymbol{k}}^s \right)
\right]^2
}{
\sum_{s, \boldsymbol{k}} f_{\boldsymbol{k}}^s \left( 1- f_{\boldsymbol{k}}^s \right)
}
.
\end{align}
Here,
$ f_{\boldsymbol{k}}^s = \{ 1 + \exp [ ( E_{0, \boldsymbol{k}}^s - E_\mathrm{F} ) / k_\mathrm{B}T ] \} ^{-1}$
is the Fermi distribution function.
$\partial E_{\boldsymbol{k}}^s \left( \varepsilon_{\Gamma_\gamma} \right) / \partial \varepsilon_{\Gamma_\gamma}
|_{ \varepsilon_{\Gamma_\gamma} \rightarrow 0} $
and
$\partial^2 E_{\boldsymbol{k}}^s \left( \varepsilon_{\Gamma_\gamma} \right) / \partial \varepsilon_{\Gamma_\gamma}^2
|_{ \varepsilon_{\Gamma_\gamma} \rightarrow 0} $
are written as
$\partial E_{\boldsymbol{k}}^s / \partial \varepsilon_{\Gamma_\gamma}$
and
$\partial^2 E_{\boldsymbol{k}}^s / \partial \varepsilon_{\Gamma_\gamma}^2$
, respectively.
The conservation law for the total electron number with respect to the strain,
$\partial n / \partial \varepsilon_{\Gamma_\gamma}
= \sum_{ \boldsymbol{k} } \partial f_{\boldsymbol{k}} / \partial \varepsilon_{\Gamma_\gamma} = 0$,
is employed to calculate Eq. (\ref{Elastic_Band Model}).
The second term on the right-hand side of Eq. (\ref{Elastic_Band Model}) corresponds to van Vleck term, which originates from the off-diagonal element $h_{\boldsymbol{k}, \Gamma_\gamma}$ in the multipole-strain interaction of Eq. (\ref{HQS_two band model}).
The third and fourth terms are the Curie terms $(\sim1/T)$ related to the diagonal elements $d_{\boldsymbol{k}, \Gamma_\gamma}^\mathrm{l}$ and $d_{\boldsymbol{k}, \Gamma_\gamma}^\mathrm{u}$.
In this two-band model, the matrix elements of a multipole and the band gap are independent on the wave vector $\boldsymbol{k}$.
The temperature dependence of the elastic constant is obtained by replacing the sum over the wave vector $\sum_{\boldsymbol{k}}$ by the energy integral using the DOS of the two-band model shown in Fig. \ref{Fig6} as
\cite{Nakamura_JMMM76},
%%%%%%%%%%%%%%%%%%%%%%%%%%%%%%%%elastic band model
\begin{align}
\label{Fitting equation}
&C_{\Gamma_\gamma}\left(T\right)
= C_{\Gamma_\gamma}^0
\nonumber \\
&-\frac{1}{4} D \left( d_{\Gamma_\gamma}^\mathrm{u} - d_{\Gamma_\gamma}^\mathrm{l} \right)^2
\left[ \tanh \left( \frac{ \Delta + W}{ 2 k_\mathrm{B} T } \right) - \tanh \left( \frac{ \Delta}{ 2 k_\mathrm{B} T } \right) \right]
\nonumber \\
& + D \left| h_{\Gamma_\gamma} \right|^2 \frac{ 2 k_\mathrm{B} T }{ \Delta }
\ln\left| \frac{ \cosh \left( \frac{ \Delta }{ 2 k_\mathrm{B} T } \right) }{ \cosh \left( \frac{ \Delta + W}{ 2 k_\mathrm{B} T } \right) } \right|
.
\end{align}
Here, we adopt the background elastic constant $C_{\Gamma_\gamma}^0 = A - B /( e^{C/T} - 1)$
\cite{Vershni_PRB2}.
$C_{\Gamma_\gamma}^0$,
$D (d_{\Gamma_\gamma}^\mathrm{u} - d_{\Gamma_\gamma}^\mathrm{l} )^2$,
and
$D | h_{\Gamma_\gamma} |^2$
in Eq. (\ref{Fitting equation}) are treated as fit parameters.
The second and third terms in Eq. (\ref{Fitting equation}) correspond to Curie and van Vleck term, respectively.

%%%%%%parameters%%%%%%%%%%%%%%%%%%%%%%%%%%
\begin{table*}
\caption{
Fit parameters determined by the analysis of the elastic constants $C_{44}$, $C_\mathrm{T}$, and $C_\mathrm{B}$ in zero field using the multipole susceptibility given in Eq. (\ref{Fitting equation}).
Parameters describing $C_\mathrm{B}$ at $58$ T are also listed.
$D = 2.25 \times 10^{27}$ K$^{-1}$m$^{-3}$ in zero field is calculated from $|d_\mathrm{B}^\mathrm{u} - d_\mathrm{B}^\mathrm{l}|  /k_\mathrm{B}$.
$|d_\mathrm{B}^\mathrm{u} - d_\mathrm{B}^\mathrm{l}|  /k_\mathrm{B}$ of $C_\mathrm{B}$ at 58 T is derived from the rigid-band approximation.
The parameters for SmB$_6$ are reproduced from Ref. \cite{Nakamura_JPSJ60_SmB6}.
}
\begin{ruledtabular}
\label{table2}
\begin{tabular}{cccccccc}
$ C_{\Gamma_\gamma}$ & $\Delta \left( \mathrm{K} \right)$ & $W \left( \mathrm{K} \right)$ & $D ( d_{\Gamma_\gamma}^\mathrm{u} - d_{\Gamma_\gamma}^\mathrm{l} )^2 \left( 10^{9} \mathrm{J}/\mathrm{m}^3 \right)$ & $A \left( 10^{10} \mathrm{J}/\mathrm{m}^3 \right)$ & $B \left( \mathrm{J}/\mathrm{m}^3 \right)$ & $C \left( \mathrm{K} \right)$ & $ |d_{\Gamma_\gamma}^\mathrm{u} - d_{\Gamma_\gamma}^\mathrm{l} | /k_\mathrm{B} \left( \mathrm{K} \right)$
\\
\hline
$C_{44}$ & 70 & 55 & 0.352 & $17.744$ & $1.25 \times 10^8$ & $32 $ & 106
\\
$C_\mathrm{T}$ & 70 & 55 & 8.58 & $11.865$ & $1.41 \times 10^9$ & $85$ & 526
\\
$C_\mathrm{B}$ $( 0\ \mathrm{T})$ & 70 & 55 & 3.30 & $20.146$ & $9.7 \times 10^6$ & $6$ & 326
\\
$C_\mathrm{B}$ $( 58\ \mathrm{T})$ & 37 & 55 & 10.3 & $20.146$ & $9.7 \times 10^6$ & $6$ & (576)
\\
$C_\mathrm{B}$ (SmB$_6$) & 160 & 150 & 25.6 & & & & 1280
\end{tabular}
\end{ruledtabular}
\end{table*}
%%%%%%%%%%%%%%%%%%%%%%%%%%%%%%%%%%%%%%

The analysis by the multipole susceptibility of Eq. (\ref{Fitting equation}) reveals the contribution of valence fluctuations to the elastic constant in zero field.
Fits to the temperature dependence of the elastic constants $C_{44}$, $C_\mathrm{T}$, and $C_\mathrm{B}$ in zero field are shown in Fig. \ref{Fig2}.
The fit parameters are summarized in Table \ref{table2}.
Here, we adopt $\Delta = 70$ K and $W = 55$ K at 0 T as determined by the analysis of specific-heat data of YbB$_{12}$ based on the rectangular two-band model
\cite{Iga_JMMM177}.
The temperature dependence of the elastic constants $C_{44}$, $C_\mathrm{T}$, and $C_\mathrm{B}$ can be well described by our model.
The energy gap $\Delta$, the bandwidth $W$, and the coefficient of the Curie term, $D ( d_{\Gamma_\gamma}^\mathrm{u} - d_{\Gamma_\gamma}^\mathrm{l} )^2$, are necessary to reproduce the additional contribution in the vicinity of $T^\star = 35$ K.
In contrast, the van Vleck contribution is not needed to explain the experimental results.
Our results indicate the importance of the multipole-strain interaction [Eq. (\ref{HQS_two band model})] to the elastic constants.
In particular, the broad increase of $C_\mathrm{B}$ below $\sim$ 40 K seems to be the result of the isotropic change of the ionic radii caused by valence fluctuations due to the $c$-$f$ hybridization.
We also tried to fit $C_\mathrm{B}$ to adopt $\Delta = 30$ K as determined by the high-field magnetoresistance
\cite{Sugiyama_JPSJ57}.
However, we are not able to reproduce the curvature change in $C_\mathrm{B}$ around 35 K (see Appendix \ref{Appendix_D}).

The multipole susceptibility also provides the renormalized multipole-strain coupling constant and the interaction anisotropy.
For the volume strain $\varepsilon_\mathrm{B}$, the first-order coefficient of the energy gap is described as
$d\Delta(\varepsilon_\mathrm{B})/d \varepsilon_\mathrm{B}|_{\varepsilon_\mathrm{B} \rightarrow 0} 
= (d_\mathrm{B}^\mathrm{u} - d_\mathrm{B}^\mathrm{l}) / 2$  from Eq. (\ref{Delta_Gap}).
We can change the variable of this relation from $\varepsilon_\mathrm{B}$ to the hydrostatic pressure $P$, because $P = C_\mathrm{B}\varepsilon_\mathrm{B}$.
In addition, we assume that $\Delta_{\boldsymbol{k}}$ in Eq. (\ref{Delta_Gap}) corresponds to the activation energy $E$ determined by resistivity measurements.
Thus, based on the hydrostatic pressure dependence of the resistivity of YbB$_{12}$
\cite{Iga_PhysB186},
we can estimate the renormalized hexadecapole-strain coupling constant $|d_\mathrm{B}^\mathrm{u} - d_\mathrm{B}^\mathrm{l}| / k_\mathrm{B}$ to be $326$ K by $dE/dP = -0.809$ K/GPa $= -8.09 \times 10^{-10}$ $\mathrm{K}/(\mathrm{J}/\mathrm{m}^3)$ for $C_\mathrm{B} = 20.146 \times 10^{10}$ J/m$^3$ (Table \ref{table2}).
This assumption also provides the DOS in zero field to be $D = 2.25 \times 10^{27}$ K$^{-1}$ m$^{-3}$ from $D ( d_\mathrm{B}^\mathrm{u} - d_\mathrm{B}^\mathrm{l} )^2 = 3.30 \times 10^{9}$ J/m$^3$ in Table \ref{table2}.
Accordingly, the coupling constant for each elastic mode was calculated (Table \ref{table2}).
The coupling constant for $C_{44}$ is approximately 5 and 3 times smaller than the coupling constant for $C_\mathrm{T}$ and $C_\mathrm{B}$, respectively.
Therefore, the dominant interaction is caused by the bulk strain with $\Gamma_1$ and the symmetry-breaking strain with $\Gamma_3$.
This result is useful to elucidate the quantum states, which carry the multipole degrees of freedom.

While valence fluctuations are caused by hexadecapole-strain interactions in YbB$_{12}$, the contribution of the fluctuations to the elasticity is unexpectedly small in zero field.
As shown in Fig. \ref{Fig2}, $C_\mathrm{B}$ does not exhibit a softening in YbB$_{12}$.
This result is quite different from the 3.8$\%$ softening in $C_\mathrm{B}$ observed for SmB$_6$.
Furthermore, the coupling constant $|d_\mathrm{B}^\mathrm{u} - d_\mathrm{B}^\mathrm{l}| / k_\mathrm{B} = 326$ K of YbB$_{12}$ is approximately 4 times smaller than $1280$ K reported for $C_\mathrm{B}$ of SmB$_6$ \cite{Nakamura_JPSJ60_SmB6}.

In contrast to zero field, strong valence fluctuations are revealed in applied magnetic fields.
A fit to the temperature-dependent data of $C_\mathrm{B}$ at 58 T is shown in Fig. \ref{Fig5} (dashed line).
The fit parameters at 58 T are also summarized in Table \ref{table2}.
In this analysis, we did not change $C_\mathrm{B}^0$ from that in zero field.
We fixed the bandwidth $W = 55 $ K as the previously proposed rigid-band model
\cite{Terashima_JPSJ86}, .
The softening with the minimum at 40 K is reproduced qualitatively.
Notably, the coefficient of the Curie term $D ( d_\mathrm{B}^\mathrm{u} - d_\mathrm{B}^\mathrm{l} )^2$ is enhanced from $3.30 \times 10^9$ J/m$^3$ at 0 T to $10.3 \times 10^9$ J/m$^3$ at 58 T.
Thus, the quantum state contributing to the Curie term of YbB$_{12}$ might approach that of SmB$_6$ in magnetic fields.
We stress that the hexadecapole-strain interaction originates from the coupling between the isotropic volume change of the crystal lattice and the change of ionic radii due to valence fluctuations.
Therefore, the larger $D ( d_\mathrm{B}^\mathrm{u} - d_\mathrm{B}^\mathrm{l} )^2$ in magnetic fields indicates the enhancement of valence fluctuations of Yb.

A reduced energy gap is a plausible result of the IM transition.
Our analysis reveals that the energy gap $2 \Delta = 140$ K at 0 T is reduced to 74 K at 58 T.
This may be attributed to the Zeeman effect that changes the energy of the $4f$ states (see Appendix \ref{Appendix_C}).
However, this two-band model cannot describe the gap closing in high fields.
Since the DOS is approximated as constant, we cannot describe the IM transition due to the overlap of the edge of DOS at the Fermi energy as schematically illustrated in Fig. \ref{Fig6}.
An analysis using more realistic DOS as proposed in a previous study
\cite{Ohashi_PRB70, Terashima_JPSJ86}
is needed to describe the field-induced metal state in high fields.

Since the DOS in zero field is estimated by using the pressure dependence of the activation energy of YbB$_{12}$, we cannot apply $D$ to estimate the coupling constant $|d_\mathrm{B}^\mathrm{u} - d_\mathrm{B}^\mathrm{l}| / k_\mathrm{B}$ in high fields.
Nevertheless, if we estimate the coupling constant assuming a field-independent rigid-band model with energy gap, the field-enhanced value of $576$ K can be obtained.
The increase in the elastic softening due to the increase in the coupling constant is also consistent with a previous theoretical study of the electron-phonon coupling mediated by conduction electrons and $f$-electrons
\cite{Rout_PhysicaB367}.
Although the model needs to be improved, this interpretation seems to be plausible.

For further discussion of the field-enhanced valence fluctuations in YbB$_{12}$, we estimated the valence change of Yb in high fields.
Previous studies on SmB$_6$ have revealed a valence change from $2.59 \pm 0.01$ at 300 K to $2.53 \pm 0.01$ at 60 K
\cite{Mizumaki_JPhysConfSer176}
and a softening of $C_\mathrm{B}$ by $3.1\%$ from 300 to 60 K
\cite{Nakamura_JPSJ60_SmB6}.
We assume that the valence change is proportional to the amount of elastic softening as a result of hexadecapole-strain interaction. 
Thus, the valence change is estimated to be -0.019 per 1$\%$ of elastic softening. 
Since the softening of $C_\mathrm{B}$ in SmB$_6$ and YbB$_{12}$ are described by the hexadecapole susceptibility based on the two-band model, we assume that the valence change per elastic softening applies to YbB$_{12}$ as well.
Since the contribution of the hexadecapole-strain interaction to the elastic softening in YbB$_{12}$, namely the coefficient of the Curie term $D ( d_\mathrm{B}^\mathrm{u} - d_\mathrm{B}^\mathrm{l} )^2$, is $2.5$ times smaller than in SmB$_6$ (Table \ref{table2}), the contribution of valence fluctuations to the elastic softening in YbB$_{12}$ is reduced by a factor of $2.5$.
At 1.4 K, in the high-field Kondo-metal phase, the 0.09\% softening from $B_\mathrm{IM}$ to 58 T (Fig. \ref{Fig3}) indicates a small valence change of only approximately $-0.00069$.
Furthermore, at 40 K, a valence change of approximately $-0.0040$ is estimated from the $0.52\%$ softening of $C_\mathrm{B}$.
Such a valence change at 40 K may be detectable by high-field synchrotron x-ray measurements
\cite{Matsuda_KPS62}.

Our results seem to be in conflict with the localized tendency of $4f$ states in the magnetic fields
\cite{Aoki_PRL71, Matsuda_PRB86, Matsuda_SCES2013}.
For a comprehensive understanding of the results, we discuss the Zeeman mixing and hybridization between Yb and B electrons in addition to the $c$-$f$ hybridization due to the $5d$ and $4f$ electrons of the Yb atoms.
In YbB$_{12}$, the contribution of the $\Gamma_6$ and $\Gamma_7$ states to the ground state is enhanced by the Zeeman effect (see appendix \ref{Appendix_C}, Fig. \ref{Fig9}).
Thus, we expect that magnetic fields reduce the anisotropy of the electronic states due to the contributions of the $\Gamma_8$, $\Gamma_6$, and $\Gamma_7$ wave functions in YbB$_{12}$.
In addition, as shown in Fig. \ref{Fig1}(a), the Yb ion of YbB$_{12}$ is surrounded by a highly isotropic cage made up of 24 borons.
This indicates an isotropic hybridization between the Yb $4f$ electrons and the B $2p$ electrons in addition to the $5d$-$4f$ hybridization.
Thus, we suggest that the valence fluctuations are induced by the interatomic $p$-$f$ hybridization due to the isotropic wave function in high fields.
Furthermore, a field-induced $p$-$f$ hybridization is consistent with the enhancement of the hexadecapole-strain interaction in high fields.
In general, the matrix element of the hexadecapole $H_0$ for the wave function $\psi$ is given by $\int d\boldsymbol{r} \psi^\ast H_0 \psi$.
Therefore, a spatially expanded wave function, which is expected due to the interatomic type $p$-$f$ hybridization, might enhance the renormalized multipole-strain coupling $d_\mathrm{B}^\mathrm{u(l)}$ in Eq. (\ref{HQS_two band model}).
Our assumption is consistent with the isotropic resistivity in the low-temperature Kondo-metal phase
\cite{Iga_JPhysConfSer200}.
Although the crystal structure and magnetic character are different from those of YbB$_{12}$, the similar mechanisms of field-induced $p$-$f$ hybridization and delocalization of $4f$ electrons have been proposed in the heavy-fermion compound CeRhIn$_5$ to describe the emergence of an anisotropic electronic state in high fields
\cite{Moll_NatComm6, Ronning_Nature548, Rosa_PRL122, Kurihara_PRB101}.

%%%%%%%%%%%%%%%%%%conclusion%%%%%%%%%%%%%%%

\section{
\label{conclusion}
Conclusion}

In the present work, we investigated valence fluctuations of YbB$_{12}$ in zero and high fields by use of ultrasonic measurements.
In zero field, the additional elastic hardening of $C_{11}$, $C_{44}$, $C_\mathrm{T} = \left(C_{11} - C_{12} \right)/2$, and the bulk modulus $C_\mathrm{B} = \left(C_{11} + 2C_{12} \right)/3$ indicates only a small contribution of valence fluctuations to the elastic constants.
In the Kondo-metal state, the valence fluctuations due to the $c$-$f$ hybridization are suggested to be enhanced by the field-induced elastic softening of $C_\mathrm{B}$.
We found signatures of strong field-induced valence fluctuations in the vicinity of 40 K.
Our phenomenological analysis of the temperature dependence of $C_\mathrm{B}$ based on the two-band model reveals that both, the additional contribution in zero field and the field-induced elastic softening, are reasonably described by the hexadecapole susceptibility.
In particular, the field-induced elastic softening is attributed to the enhancement of the hexadecapole-strain coupling.
This result indicates that the magnetic field enhances an isotropic volume change of the crystal lattice and the change of ionic radii due to valence fluctuations.
Therefore, we propose field-induced valence fluctuations due to $c$-$f$ hybridization in YbB$_{12}$.
In particular, we propose that the $p$-$f$ hybridization between Yb-$4f$ and B-$2p$ electrons plays a key role in high fields.
The observed decrease of the energy gap in magnetic fields is explained by 
the energy shift of the $4f$ electrons due to the Zeeman effect.

Our study shows that ultrasonic measurements are useful to detect valence fluctuations.
As suggested by a theoretical work
\cite{Watanabe_JPSJ89},
such measurements may play a key role in the study of valence quantum criticality.
We expect that field-induced valence fluctuations appear in other valence-fluctuating compounds.

%%%%%%%%%%%%%%%%%%Acknowledgement%%%%%%%%%%%%%%%%%%%%
\section*{Acknowledgment}
The authors thank Yuichi Nemoto and Mitsuhiro Akatsu for supplying the LiNbO$_3$ piezoelectric transducers.
We also thank Keisuke Mitsumoto and Shintaro Nakamura for valuable discussions.
This work was partly supported by JSPS Bilateral Joint Research Projects (JPJSBP120193507) and Grants-in-Aid for young scientists (KAKENHI JP20K14404). We acknowledge the support of the HLD at HZDR, member of the European Magnetic Field Laboratory (EMFL), the Deutsche Forschungsgemeinschaft (DFG) through the W\"urzburg-Dresden Cluster of Excellence on Complexity and Topology in Quantum Matter $-$ $ct.qmat$ (EXC 2147, project No. 390858490), and the BMBF via DAAD (project No 57457940).

 %%%%%%%%%%%%%%%%%%Appendix%%%%%%%%%%%%%%%
\appendix

\section{
\label{Appendix_A}
Temperature dependence of elastic constant $C_{11}$
}

\begin{figure}[htbp]
\begin{center}
\includegraphics[clip, width=0.5\textwidth, bb=0 0 470 284]{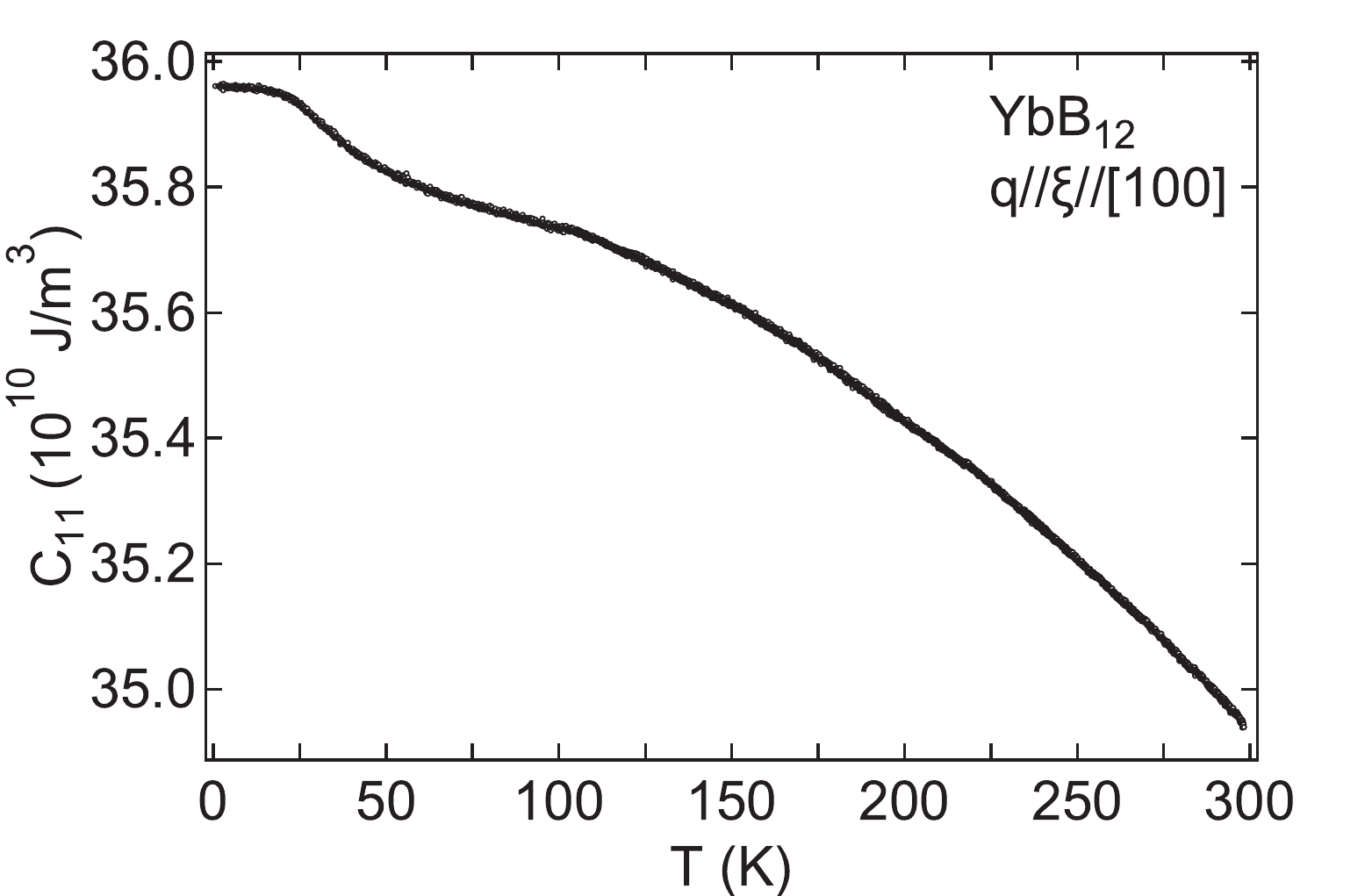}
\end{center}
\caption{
Temperature dependence of the longitudinal elastic constant $C_{11}$.
}
\label{Fig7}
\end{figure}

Figure \ref{Fig7} shows the temperature dependence of the elastic constant $C_{11}$ in a wide temperature range of up to 300 K.
$C_{11}$ exhibits an increase with decreasing temperatures from 300 K.
This result indicates the elastic hardening of the bulk modulus $C_\mathrm{B}$ from 300 K down to low temperatures.

%%%%%%%%%%%%%%%%%%%%%%%%Fig7%%%%%%%%%
%%%%%%%%%%%%%%%%%%%%%%%%%%%%%%%%%%%%%

\section{
\label{Appendix_B}
Multipole susceptibility for CEF wave functions in zero field
}

Here, we present the CEF wave functions, the multipole matrices, and multipole susceptibility of YbB$_{12}$ assuming the localized $4f$ electrons.
We show that the multipole susceptibility cannot describe our experimental results in Fig. \ref{Fig2}. 

To calculate the multipole susceptibility of YbB$_{12}$, we use CEF wave functions of the $4f$ electrons for Yb$^{3+}$ with the total angular momentum $J = 7/2$.
The CEF Hamiltonian $H_\mathrm{CEF}$ under $O_h$ symmetry is written as
%CEF Hamiltonian
\begin{align}
\label{H_CEF}
H_\mathrm{CEF}
= B_4 \left( O_4^0 + 5 O_4^4\right) + B_6 \left( O_6^0 -21 O_6^4\right)
.
\end{align}
Here, $B_4$ and $B_6$ are the CEF parameters.
The matrix elements of $O_4^0$, $O_4^4$, $O_6^0$, and $O_6^4$ for $\left| J_z \right\rangle$ are listed in Ref. \cite{Hutchings}.
The wave functions diagonalizing $H_\mathrm{CEF}$ are given by
\cite{Kanai_JPSJ84}
%CEF wavefunctions
\begin{align}
\label{CEF _G81}
\left| \Gamma_8^{1\pm} \right\rangle
&= -\sqrt{ \frac{7}{12} } \left| \pm \frac{7}{2} \right\rangle + \sqrt{ \frac{5}{12} } \left|  \mp \frac{1}{2} \right\rangle,
\\
\label{CEF_G82}
\left| \Gamma_8^{2\pm} \right\rangle
&= \frac{1}{2} \left| \pm \frac{5}{2} \right\rangle + \frac{\sqrt{3}}{2} \left|  \mp \frac{3}{2} \right\rangle,
\\
\label{CEF_G6}
\left| \Gamma_6^{1 \pm} \right\rangle
&= \sqrt{ \frac{5}{12} } \left| \pm \frac{7}{2} \right\rangle + \sqrt{ \frac{7}{12} } \left|  \mp \frac{1}{2} \right\rangle,
\\
\label{CEF_G7}
\left| \Gamma_7^\pm \right\rangle
&= -\frac{\sqrt{3}}{2} \left| \pm \frac{5}{2} \right\rangle + \frac{1}{2} \left|  \mp \frac{3}{2} \right\rangle
,
\end{align}
where
$\left| \Gamma_8^{1\pm} \right>$ and $\left| \Gamma_8^{2\pm} \right>$
are the ground-state wave functions and 
$\left| \Gamma_7^\pm \right>$ and $\left| \Gamma_6^\pm \right>$
are the degenerate excited states.
The matrix elements of $H_\mathrm{CEF}$ for the wave functions given in Eqs. (\ref{CEF _G81}) - (\ref{CEF_G7}) provide the eigenenergy of each CEF state described as
$E_{\Gamma_8} = 120\left( B_4 + 168B_6\right)$,
$E_{\Gamma_6} = 120\left( 7B_4 - 210B_6\right)$,
and 
$E_{\Gamma_7} = -120\left( 9B_4 + 126B_6\right)$.
The energy gap $\Delta_\mathrm{CEF} = 23$ meV = $270$ K between the ground state and the excited states $\Gamma_6$ and $\Gamma_7$
\cite{Kanai_JPSJ84}
provides the CEF parameters $B_4 = -33.7$ meV and $B_6 = -6.24$ meV.

\begin{figure}[htbp]
\begin{center}
\includegraphics[clip, width=0.5\textwidth, bb=0 0 390 480]{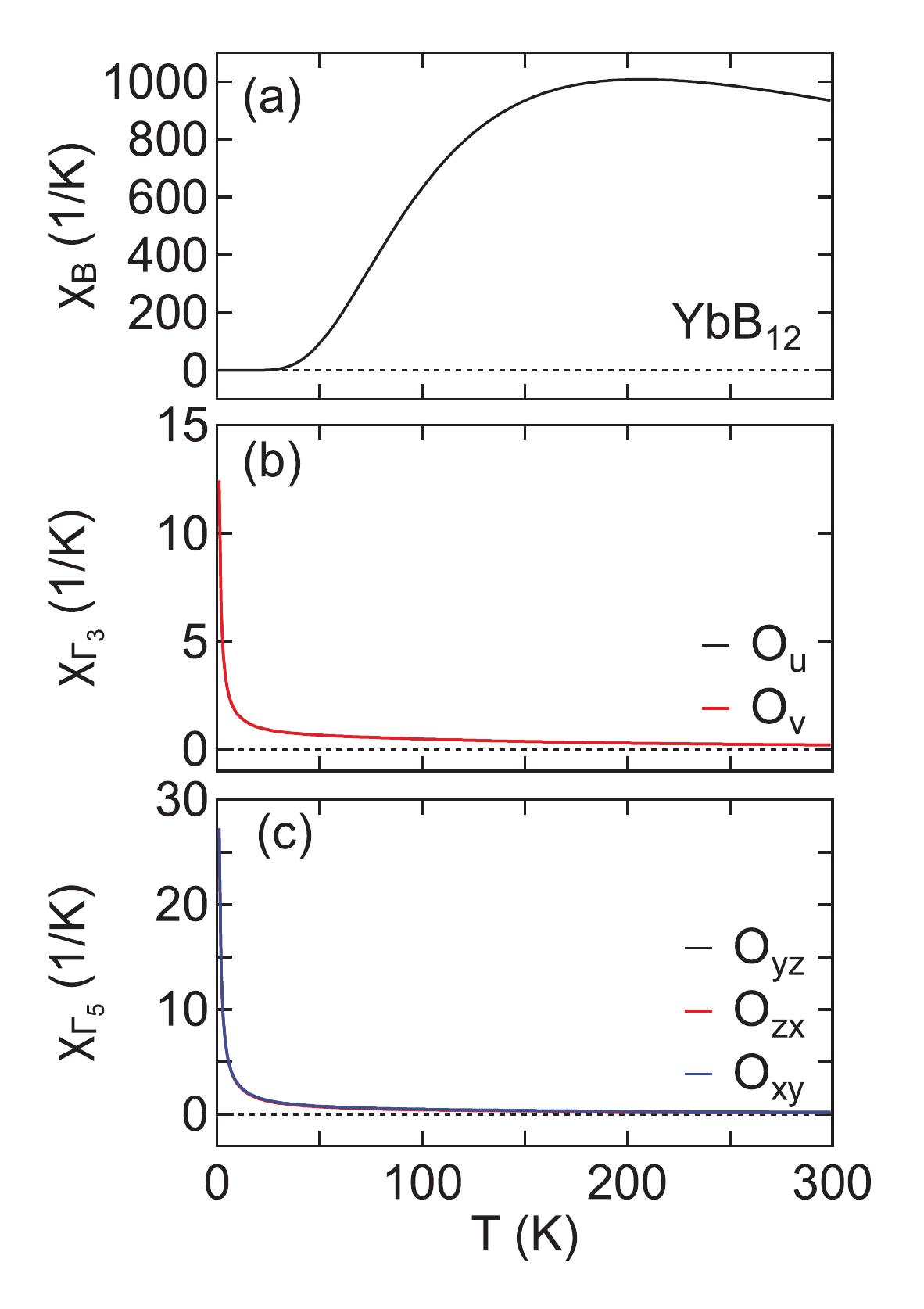}
\end{center}
\caption{
Temperature dependence of the electric multipole susceptibility of YbB$_{12}$.
(a) Hexadecapole susceptibility $\chi_\mathrm{B}$ of $H_0 = O_4^0 + 5 O_4^4$ with $\Gamma_1$.
This susceptibility provides the temperature dependence of the bulk modulus $C_\mathrm{B}$.
(b) Quadrupole susceptibility $\chi_{\Gamma_3}$ of $O_u$ and $O_v$ with $\Gamma_3$ related to $(C_{11} - C_{12})/2$
(c) Quadrupole susceptibility $\chi_{\Gamma_5}$ of $O_{yz}$, $O_{zx}$, and $O_{xy}$ with $\Gamma_5$ related to $C_{44}$.
As indicated in Eq. (\ref{elastic}),  $-\chi_{\Gamma_\gamma}$ contributes to these elastic constants. (The curves for $O_u$ and $O_v$ as well as for $O_{yz}$, $O_{zx}$, and $O_{xy}$ virtually lie on top of each other).
}
\label{Fig8}
\end{figure}

The matrices of the hexadecapole $H_0 = O_4^0 + 5 O_4^4$ with irrep $\Gamma_1$, the quadrupoles $O_{u}$ and $O_{v}$ with $\Gamma_3$, and $O_{yz}$, $O_{zx}$,  and $O_{xy}$ with $\Gamma_5$ for the wave functions (\ref{CEF _G81}) - (\ref{CEF_G7}) are calculated as
%matrices
\begin{widetext}
\begin{align}
\label{H_0}
H_0
&= \bordermatrix{
& \left| \Gamma_8^{1+} \right\rangle & \left| \Gamma_8^{1-} \right\rangle & \left| \Gamma_8^{2+} \right\rangle & \left| \Gamma_8^{2-} \right\rangle & \left| \Gamma_6^+ \right\rangle & \left| \Gamma_6^- \right\rangle & \left| \Gamma_7^+ \right\rangle & \left| \Gamma_7^- \right\rangle
\cr
& 120 & 0 & 0 & 0 & 0 & 0 & 0 & 0
\cr
& 0 & 120 & 0 & 0 & 0 & 0 & 0 & 0
\cr
& 0 & 0 & 120 & 0 & 0 & 0 & 0 & 0
\cr
& 0 & 0 & 0 & 120 & 0 & 0 & 0 & 0 
\cr
& 0 & 0 & 0 & 0 & 840 & 0 & 0 & 0
\cr
& 0 & 0 & 0 & 0 & 0 & 840 & 0 & 0 
\cr
& 0 & 0 & 0 & 0 & 0 & 0 & 1080 & 0 
\cr
& 0 & 0 & 0 & 0 & 0 & 0 & 0 & 1080 
},
\end{align}
\begin{align}
\label{O_u}
O_{u}
&= \begin{pmatrix}
6 & 0 & 0 & 0 & -3\sqrt{35} & 0 & 0 & 0
\cr
0 & 6 & 0 & 0 & 0 & -3\sqrt{35} & 0 & 0
\cr
0 & 0 & -6 & 0 & 0 & 0 & -3\sqrt{3} & 0
\cr
0 & 0 & 0 & -6 & 0 & 0 & 0 & -3\sqrt{3} 
\cr
-3\sqrt{35} & 0 & 0 & 0 & 0 & 0 & 0 & 0
\cr
0 & -3\sqrt{35} & 0 & 0 & 0 & 0 & 0 & 0 
\cr
0 & 0 & -3\sqrt{3} & 0 & 0 & 0 & 0 & 0 
\cr
0 & 0 & 0 & -3\sqrt{3} & 0 & 0 & 0 & 0 
\end{pmatrix},
\end{align}
\begin{align}
\label{O_v}
O_{v} 
&= \begin{pmatrix}
0 & 0 & 0 & \sqrt{105} & 0 & 0 & 0 & 0
\cr
0 & 0 & 2\sqrt{3} & 0 & 0 & 0 & -3 & 0
\cr
0 & 2\sqrt{3} & 0 & 0 & 0 & \sqrt{105} & 0 & 0
\cr
\sqrt{105} & 0 & 0 & 0 & 2\sqrt{3} & 0 & 0 & 0 
\cr
0 & 0 & 0 & 2\sqrt{3} & 0 & 0 & 0 & -3
\cr
0 & 0 & \sqrt{105} & 0 & 0 & 0 & 0 & 0 
\cr
0 & -3 & 0 & 0 & 0 & 0 & 0 & 0 
\cr
0 & 0 & 0 & 0 & -3 & 0 & 0 & 0 
\end{pmatrix},
\end{align}
\begin{align}
\label{O_yz}
O_{yz}
& = \begin{pmatrix}
0 & 0 & 0 & 0 & 0 & 0 & \sqrt{35}i & 0
\cr
0 & 0 & 0 & 3\sqrt{3}i & 0 & 0 & 0 & -4i
\cr
0 & 0 & 0 & 0 & -3\sqrt{3}i & 0 & 0 & -2\sqrt{3}i
\cr
0 & -3\sqrt{3}i & 0 & 0 & 0 & 0 & -2\sqrt{3}i & 0 
\cr
0 & 0 & 3\sqrt{3}i & 0 & 0 & 0 & -4i & 0
\cr
0 & 0 & 0 & 0 & 0 & 0 & 0 & \sqrt{35}i 
\cr
-\sqrt{35}i & 0 & 0 & 2\sqrt{3}i & 4i & 0 & 0 & 0 
\cr
0 & 4i & 2\sqrt{3}i & 0 & 0 & -\sqrt{35}i & 0 & 0 
\end{pmatrix},
\end{align}
\begin{align}
\label{O_zx}
O_{zx} 
&= \begin{pmatrix}
0 & 0 & 0 & 0 & 0 & 0 & -\sqrt{35} & 0
\cr
0 & 0 & 0 & 3\sqrt{3} & 0 & 0 & 0 & -4
\cr
0 & 0 & 0 & 0 & -3\sqrt{3} & 0 & 0 & 4\sqrt{3}
\cr
0 & 3\sqrt{3} & 0 & 0 & 0 & 0 & -4\sqrt{3} & 0 
\cr
0 & 0 & -3\sqrt{3} & 0 & 0 & 0 & 4 & 0
\cr
0 & 0 & 0 & 0 & 0 & 0 & 0 & \sqrt{35} 
\cr
-\sqrt{35} & 0 & 0 & -4\sqrt{3} & 4 & 0 & 0 & 0 
\cr
0 & -4 & 4\sqrt{3} & 0 & 0 & \sqrt{35} & 0 & 0 
\end{pmatrix},
\end{align}
\begin{align}
\label{O_xy}
O_{xy}
&= \begin{pmatrix}
0 & 0 & 0 & 0 & 0 & 0 & 0 & -\sqrt{35}i
\cr
0 & 0 & 3\sqrt{3}i & 0 & 0 & 0 & 8i & 0
\cr
0 & -3\sqrt{3}i & 0 & 0 & 0 & 0 & 0 & 0
\cr
0 & 0 & 0 & 0 & 3\sqrt{3}i & 0 & 0 & 0 
\cr
0 & 0 & 0 & -3\sqrt{3}i & 0 & 0 & 0 & -8i
\cr
0 & 0 & 0 & 0 & 0 & 0 & \sqrt{35}i & 0 
\cr
0 & -8i & 0 & 0 & 0 & -\sqrt{35}i & 0 & 0 
\cr
\sqrt{35}i & 0 & 0 & 0 & 8i & 0 & 0 & 0 
\end{pmatrix}.
\end{align}
\end{widetext}
Here, Stevens equivalent operators 
$O_{u} = 3J_z^2 - J\left( J + 1 \right)$,
$O_{v} = J_x^2 - J_y^2$,
$O_{yz} = J_yJ_z + J_zJ_y$,
$O_{zx} = J_zJ_x + J_xJ_x$,
and
$O_{xy} = J_xJ_y + J_yJ_x$,
given by the components of the total angular momentum $J_x$, $J_y$, and $J_z$, are used to calculate the matrix elements.
Considering the second-order perturbation processes for the $i$-th CEF state with energy $E_i^0$ due to the multipole-strain interaction of Eq. (\ref{H_MS}), which is described as 
\begin{align}
%perturbation
\label{perturb}
E_i \left( \varepsilon_{\Gamma_\gamma} \right)
= E_i^0 &- g_{\Gamma_\gamma} \langle i \left| O_{\Gamma_\gamma} \right| i \rangle \varepsilon_{\Gamma_\gamma}
\nonumber \\
&- g_{\Gamma_\gamma}^2 \sum_{j \neq i}
\frac{ \left| \langle i \left| O_{\Gamma_\gamma} \right| j \rangle \right|^2} 
{E_j^0 - E_i^0} \varepsilon_{\Gamma_\gamma}^2
,
\end{align}
the total free energy $F$, that consists of the CEF state and the strain, is written as
%Free energy
\begin{align}
\label{Free energy}
F \left(T, \varepsilon_{\Gamma_\gamma} \right)
= \frac{1}{2} C_{\Gamma_\gamma}^0 \varepsilon_{\Gamma_\gamma}^2
- N k_\mathrm{B} T \ln Z\left( \varepsilon_{\Gamma_\gamma} \right)
.
\end{align}
Here, $N$ is the number of Yb ions per unit volume and 
$Z ( \varepsilon_{\Gamma_\gamma} )$ is the partition function written as
$Z( \varepsilon_{\Gamma_\gamma} )
= \sum_i \exp\left[ - E_i ( \varepsilon_{\Gamma_\gamma} )/k_\mathrm{B}T \right]$.
Thus, the elastic constant and the multipole susceptibility are calculated as
%susceptibility
\begin{align}
\label{elastic}
&C_{\Gamma_\gamma}
= \frac{ \partial^2 F }{ \partial \varepsilon_{\Gamma_\gamma}^2 }
= C_{\Gamma_\gamma}^0 
- Ng_{\Gamma_\gamma}^2 \chi_{\Gamma_\gamma}
, \\
\label{suscep}
&-g_{\Gamma_\gamma}^2 \chi_{\Gamma_\gamma}
= \left\langle \frac{\partial^2 E  }{\partial \varepsilon_{\Gamma_\gamma}^2 } \right\rangle
\nonumber \\
&-\frac{1}{k_\mathrm{B} T}
\left\{
\left\langle \left( \frac{\partial E  }{\partial \varepsilon_{\Gamma_\gamma} } \right)^2 \right\rangle
- \left\langle \frac{\partial E  }{\partial \varepsilon_{\Gamma_\gamma} } \right\rangle^2
\right\}
.
\end{align}
Here, $C_{\Gamma_\gamma}^0$ is a background elastic constant, $\langle A \rangle$ is the thermal average using Boltzmann statistics written as
$\langle A \rangle
= \sum_i A_i \exp[-E_i/k_\mathrm{B} T]/Z$, 
and
$\partial E ( \varepsilon_{\Gamma_\gamma} ) / \partial \varepsilon_{\Gamma_\gamma}
|_{ \varepsilon_{\Gamma_\gamma} \rightarrow 0} $
and
$\partial^2 E ( \varepsilon_{\Gamma_\gamma} ) / \partial \varepsilon_{\Gamma_\gamma}^2
|_{ \varepsilon_{\Gamma_\gamma} \rightarrow 0} $
are written as 
$\partial E / \partial \varepsilon_{\Gamma_\gamma}$
and
$\partial^2 E / \partial \varepsilon_{\Gamma_\gamma}^2$, respectively.
The first term on the right-hand side of Eq. (\ref{suscep}) corresponds to van Vleck term being constant at low temperatures and the second one to Curie term showing the reciprocal temperature dependence.  
The calculated multipole susceptibility $\chi_{\Gamma_\gamma}$ is shown in Fig. \ref{Fig8}.

The hexadecapole susceptibility $\chi_\mathrm{B}$ would indicate a monotonic hardening of $C_\mathrm{B}$ below 100 K down to low temperatures because the temperature dependence of the elastic constant $C_\mathrm{\Gamma_\gamma}$ is given by  $-\chi_\mathrm{B}$, i.e., the second term in Eq. (\ref{elastic}).
The divergent behavior of $\chi_{\Gamma_3}$ and $\chi_{\Gamma_5}$ would predict an elastic softening of $(C_{11} - C_{12})/2$ and $C_{44}$ at low temperatures, respectively.
However, our experimental results of YbB$_{12}$ in zero field cannot be described by the susceptibility based on CEF wave functions using this picture, i.e., localized $4f$ electrons.
%%%%%%%%%%%%%%%%%%%%%%%%Fig8%%%%%%%%%
%%%%%%%%%%%%%%%%%%%%%%%%%%%%%%%%%%%%%%%%%%%%%%%%%%%%%%%%%%%%%

\section{
\label{Appendix_C}
Hexadecapole susceptibility for CEF wave functions in magnetic fields
}

In this section, the CEF wave functions, the hexadecapole matrix, and the hexadecapole susceptibility in magnetic fields of YbB$_{12}$ are presented assuming localized $4f$ electrons.
We show that the elastic softening of $C_\mathrm{B}$ in high fields cannot be described by the hexadecapole susceptibility $\chi_\mathrm{B}$.

To calculate the hexadecapole susceptibility in magnetic fields, we consider the Zeeman Hamiltonian for $B\|[001]$ given by
%Zeeman Hamiltonian
\begin{align}
\label{H_Zeeman}
H_\mathrm{Zeeman}
=-g_J \mu_\mathrm{B}B_0J_z
.
\end{align}
Here, $g_J$ is the Land\'e $g$-factor, $\mu_\mathrm{B}$ is the Bohr magneton, $B_0$ is the magnetic field, and $J_z$ is the magnetic dipole.
Using the CEF wave functions of Eqs. (\ref{CEF _G81})-(\ref{CEF_G7}), the matrix of $H_\mathrm{Zeeman}$ of Eq. (\ref{H_Zeeman})  is written as
%Matrix of H_Zeeman
\begin{widetext}
\begin{align}
\label{Matrix of H_Zeeman}
H_\mathrm{Zeeman}
&= \begin{pmatrix}
-\frac{11}{6}B & 0 & 0 & 0 & \frac{\sqrt{35}}{3}B & 0 & 0 & 0
\cr
0 & \frac{11}{6}B & 0 & 0 & 0 & -\frac{\sqrt{35}}{3}B & 0 & 0
\cr
0 & 0 & \frac{1}{2}B & 0 & 0 & 0 & \sqrt{3}B & 0
\cr
0 & 0 & 0 & -\frac{1}{2}B & 0 & 0 & 0 & -\sqrt{3}B 
\cr
\frac{\sqrt{35}}{3}B & 0 & 0 & 0 & -\frac{7}{6}B & 0 & 0 & 0
\cr
0 & -\frac{\sqrt{35}}{3}B & 0 & 0 & 0 & \frac{7}{6}B & 0 & 0 
\cr
0 & 0 & \sqrt{3}B & 0 & 0 & 0 & -\frac{3}{2}B & 0 
\cr
0 & 0 & 0 & -\sqrt{3}B & 0 & 0 & 0 & \frac{3}{2}B
\end{pmatrix}.
\end{align}
Here, for the convenience, $B$ in the matrix elements of Eq. (\ref{Matrix of H_Zeeman}) is set as $B = g_J \mu_\mathrm{B}B_0$.
The total Hamiltonian $H_\mathrm{total} = H_\mathrm{CEF} + H_\mathrm{Zeeman}$ is diagonalized as
%Matrix of H_Zeeman
\begin{align}
\label{Matrix of H_field}
H_\mathrm{total}
= \bordermatrix{
& \left| 1+ \right\rangle & \left| 1- \right\rangle & \left| 2+ \right\rangle & \left| 2- \right\rangle & \left| 3+ \right\rangle & \left| 3- \right\rangle & \left| 4+ \right\rangle & \left| 4- \right\rangle
\cr
& E_1^+ & 0 & 0 & 0 & 0 & 0 & 0 & 0
\cr
& 0 & E_1^- & 0 & 0 & 0 & 0 & 0 & 0
\cr
& 0 & 0 & E_2^+ & 0 & 0 & 0 & 0 & 0
\cr
& 0 & 0 & 0 & E_2^- & 0 & 0 & 0 & 0 
\cr
& 0 & 0 & 0 & 0 & E_3^+ & 0 & 0 & 0
\cr
& 0 & 0 & 0 & 0 & 0 & E_3^- & 0 & 0 
\cr
& 0 & 0 & 0 & 0 & 0 & 0 & E_4^+ & 0 
\cr
& 0 & 0 & 0 & 0 & 0 & 0 & 0 & E_4^- 
},
\end{align}
\end{widetext}
Here, the eigen energis in the matrix of Eq. (\ref{Matrix of H_field}) are written as
\begin{align}
%%%%%%Eiven value
\label{E_1pm}
E_1^\pm
= \frac{ \Delta_\mathrm{CEF} - 3 B}{2} \pm \delta E_1
,
\end{align}
\begin{align}
\label{E_2pm}
E_2^\pm
= \frac{ \Delta_\mathrm{CEF} + 3 B}{2} \pm \delta E_2
,
\end{align}
\begin{align}
\label{E_3pm}
E_3^\pm
= \frac{ \Delta_\mathrm{CEF} - B}{2} \pm \delta E_3
,
\end{align}
\begin{align}
\label{E_4pm}
E_4^\pm
= \frac{ \Delta_\mathrm{CEF} + B}{2} \pm \delta E_4
.
\end{align}
For convenience, $\delta E_i$ in Eqs. (\ref{E_1pm}) - (\ref{E_4pm}) is set as
%%%%%%%%delta E
\begin{align}
\label{dE_1}
\delta E_1
= \sqrt{ \Delta E_1^2 + \frac{35}{9}B^2 }
,
\end{align}
\begin{align}
\label{dE_2}
\delta E_2
= \sqrt{ \Delta E_2^2 +  \frac{35}{9}B^2 }
,
\end{align}
\begin{align}
\label{dE_3}
\delta E_3
= \sqrt{ \Delta E_3^2 + 3B^2 }
,
\end{align}
\begin{align}
\label{dE_4}
\delta E_4
= \sqrt{ \Delta E_4^2 + 3B^2 }
.
\end{align}
We also set $\Delta E_i$ in Eqs. (\ref{dE_1}) - (\ref{dE_4}) as follows:
%%%%%%%%DeltaE
\begin{align}
\label{DeltaE_1}
\Delta E_1 = \frac{ \Delta_\mathrm{CEF} }{2} + \frac{1}{3}B
, 
\end{align}
\begin{align}
\label{DeltaE_2}
\Delta E_2 = \frac{ \Delta_\mathrm{CEF} }{2} - \frac{1}{3}B
, 
\end{align}
\begin{align}
\label{DeltaE_3}
\Delta E_3 = \frac{ \Delta_\mathrm{CEF} }{2} - B
, 
\end{align}
\begin{align}
\label{DeltaE_4}
\Delta E_4 = \frac{ \Delta_\mathrm{CEF} }{2} + B
.
\end{align}
%%%%%%%%%%%%%%%%%%%%%%%%Fig9%%%%%%%%%
\begin{figure}[t]
\begin{center}
\includegraphics[clip, width=0.5\textwidth, bb=0 0 500 550]{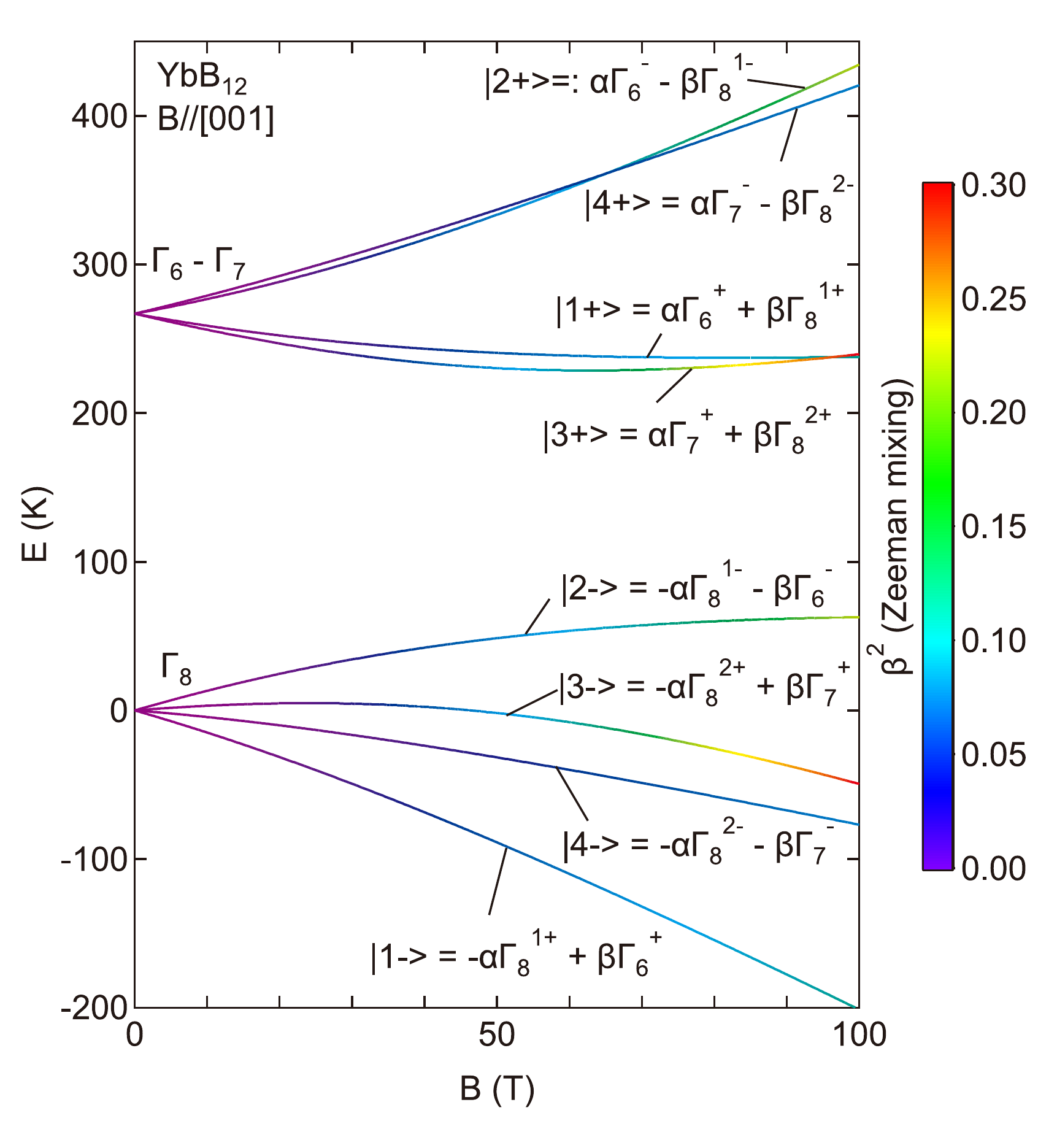}
\end{center}
\caption{
Magnetic-field dependence of the eigenenergy in $H_\mathrm{total}$ of Eq. (\ref{Matrix of H_field}) for $B\|[001]$.
The color code shows the Zeeman mixing ratio $\beta_i^2$ in the wave functions of Eqs. (\ref{WF1+})-(\ref{WF4-}).
}
\label{Fig9}
\end{figure}
%%%%%%%%%%%%%%%%%%%%%%%%%%%%%%%%%%%%%%%%%%%%%%%%%%%%%%%%%%%%%
%%%%%%%%%%%%%%%%%%%%%%%%Fig10%%%%%%%%%
\begin{figure}[htb]
\begin{center}
\includegraphics[clip, width=0.4\textwidth, bb=0 0 400 700]{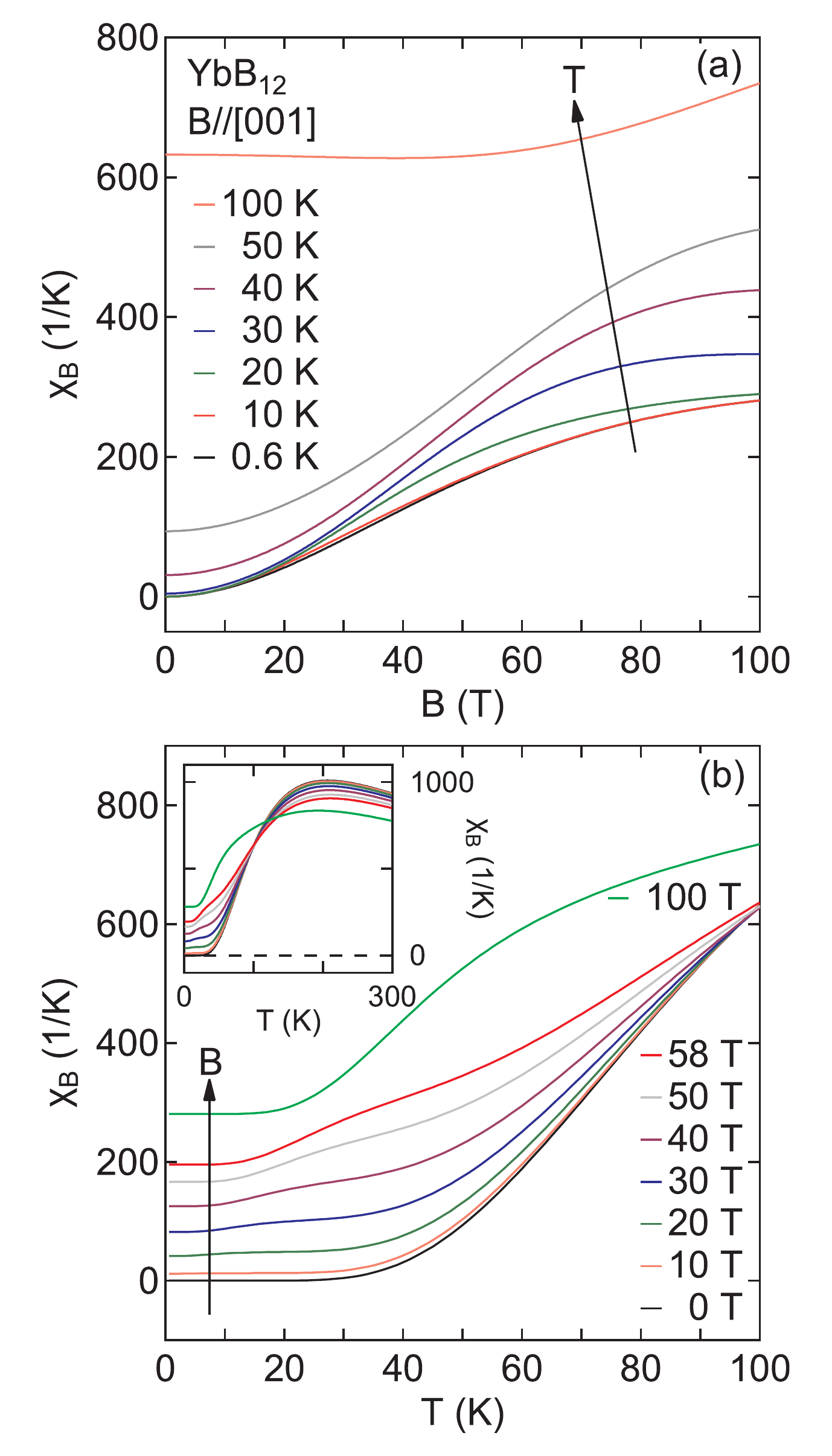}
\end{center}
\caption{
Electric hexadecapole susceptibility $\chi_\mathrm{B}$ of YbB$_{12}$ for $B\|[001]$.
(a) Magnetic-field dependence of $\chi_\mathrm{B}$ at several temperatures.
(b) Temperature dependence of $\chi_\mathrm{B}$ below 100 K in several magnetic fields. 
In the inset in (b), $\chi_\mathrm{B}$ is shown up to 300 K.
}
\label{Fig10}
\end{figure}
%%%%%%%%%%%%%%%%%%%%%%%%%%%%%%%%%%%%%%%%%%%%%%%%%%%%%%%%%%%%%
The wave functions diagonalizing the matrix Eq. (\ref{Matrix of H_field}) are written as
%%%%%%wave fucntions in fields
\begin{align}
\label{WF1+}
\left| 1 + \right\rangle
= \alpha_1 \left| \Gamma_6^+ \right\rangle + \beta_1 \left| \Gamma_8^{1+} \right\rangle
, 
\end{align}
\begin{align}
\label{WF1-}
\left| 1 - \right\rangle
= \beta_1 \left| \Gamma_6^+ \right\rangle - \alpha_1  \left| \Gamma_8^{1+} \right\rangle
, 
\end{align}
\begin{align}
\label{WF2+}
\left| 2 + \right\rangle
= \alpha_2 \left| \Gamma_6^- \right\rangle - \beta_2 \left| \Gamma_8^{1-} \right\rangle
, 
\end{align}
\begin{align}
\label{WF2-}
\left| 2 - \right\rangle
= -\beta_2 \left| \Gamma_6^- \right\rangle - \alpha_2  \left| \Gamma_8^{1-} \right\rangle
,
\end{align}
\begin{align}
\label{WF3+}
\left| 3 + \right\rangle
= \alpha_3 \left| \Gamma_7^+ \right\rangle + \beta_3 \left| \Gamma_8^{2+} \right\rangle
, 
\end{align}
\begin{align}
\label{WF3-}
\left| 3 - \right\rangle
= \beta_3 \left| \Gamma_7^+ \right\rangle - \alpha_3  \left| \Gamma_8^{2+} \right\rangle
, 
\end{align}
\begin{align}
\label{WF4+}
\left| 4 + \right\rangle
= \alpha_4 \left| \Gamma_7^- \right\rangle - \beta_4 \left| \Gamma_8^{2-} \right\rangle
, 
\end{align}
\begin{align}
\label{WF4-}
\left| 4 - \right\rangle
= -\beta_4 \left| \Gamma_7^- \right\rangle - \alpha_4  \left| \Gamma_8^{2-} \right\rangle
.
\end{align}
Here, the coefficients $\alpha_i$ and $\beta_i$ for $i = 1, 2, 3, 4$ in each wave function in Eqs. (\ref{WF1+}) - (\ref{WF4-}) are set as
%%%%%%%%%%WF_coeficients
\begin{align}
\label{alpha_1}
\alpha_1
&= \frac{\Delta E_1 + \delta E_1}{ \sqrt{ \left( \Delta E_1 + \delta E_1 \right)^2 + \frac{35}{9} B^2 } } 
, 
\end{align}
\begin{align}
\label{alpha_2}
\alpha_2
&= \frac{\Delta E_2 + \delta E_2}{ \sqrt{ \left( \Delta E_2 + \delta E_2 \right)^2 + \frac{35}{9} B^2 } } 
, 
\end{align}
\begin{align}
\label{alpha_3}
\alpha_3
&= \frac{\Delta E_3 + \delta E_3}{ \sqrt{ \left( \Delta E_3 + \delta E_3 \right)^2 + 3B^2 } } 
, 
\end{align}
\begin{align}
\label{alpha_4}
\alpha_4
&= \frac{\Delta E_4 + \delta E_4}{ \sqrt{ \left( \Delta E_4 + \delta E_4 \right)^2 + 3B^2 } } 
, 
\end{align}
\begin{align}
\label{beta}
\beta_i
&= \sqrt{ 1 - \alpha_i^2 }
.
\end{align}
The magnetic-field dependence of the eigenenergies $E_i^\pm$ of Eqs. (\ref{E_1pm})-(\ref{E_4pm}) are shown in Fig. \ref{Fig9}.
This result is consistent with the previous calculation for YbB$_{12}$
\cite{Terashima_JPSJ86}.
The multipole susceptibility of Eq. (\ref{suscep}) in magnetic field is calculated using the wave functions of Eqs. (\ref{WF1+}) - (\ref{WF4-}), the energy of Eqs. (\ref{E_1pm}) - (\ref{E_4pm}), the multipole matrices of Eqs. (\ref{H_0}) - (\ref{O_xy}), the second-order perturbation of Eq. (\ref{perturb}), and the free energy of Eq. (\ref{Free energy}).

In particular, we show the field-dependent hexadecapole susceptibility of $H_0$ in Fig. \ref{Fig10}.
Here, the matrix of the hexadecapole $H_0$ is written as
%Matrix of H_0
\begin{widetext}
\begin{align}
\label{Matrix of H_0}
H_0
&= \bordermatrix{
& \left| 1+ \right\rangle & \left| 1- \right\rangle 
\cr
& 120\left( 6\alpha_1^2 + 1 \right) & 720\alpha_1 \beta_1
\cr
& 720\alpha_1 \beta_1 & -120\left( 6\alpha_1^2 - 7 \right) 
}
\oplus
\bordermatrix{
& \left| 2+ \right\rangle & \left| 2- \right\rangle 
\cr
& 120\left( 6\alpha_2^2 + 1 \right) & -720\alpha_2 \beta_2
\cr
& -720\alpha_2 \beta_2 & -120\left( 6\alpha_2^2 - 7 \right) 
}
\nonumber \\
&\oplus
\bordermatrix{
& \left| 3+ \right\rangle & \left| 3- \right\rangle 
\cr
& -120\left( 10\alpha_3^2 - 1 \right) & -1200\alpha_3 \beta_3
\cr
& -1200\alpha_3 \beta_3 & 120\left( 10\alpha_3^2 - 9 \right)
}
\oplus
\bordermatrix{
& \left| 4+ \right\rangle & \left| 4- \right\rangle 
\cr
& -120\left( 10\alpha_4^2 - 1 \right) & -1200\alpha_4 \beta_4
\cr
& -1200\alpha_4 \beta_4 & 120\left( 10\alpha_4^2 - 9 \right)
}
.
\end{align}
\end{widetext}
The experimental results of the magnetic-field dependence of $C_\mathrm{B}$ at 20, 40, and 50 K (Fig. \ref{Fig3}) can be qualitatively reproduced by the hexadecapole susceptibility $\chi_\mathrm{B}$ shown in Fig. \ref{Fig10}(a).
However, the experimental results of the elastic softening of $C_\mathrm{B}$ in high fields (Fig. \ref{Fig5}) cannot be described by $\chi_\mathrm{B}$ shown in Fig. \ref{Fig10}(b) since $\chi_\mathrm{B}$ indicates a hardening of $C_\mathrm{B}$ towards lower temperatures.
Therefore, our experimental results of YbB$_{12}$ in high magnetic fields cannot be described by the susceptibility based on CEF wave functions of localized $4f$ electrons.

\section{
\label{Appendix_D}
Hexadecapole susceptibility for smaller energy gap
}

%%%%%%%%%%%%%%%%%%%%%%%%Fig11%%%%%%%%%
\begin{figure}[htbp]
\begin{center}
\includegraphics[clip, width=0.5\textwidth, bb=0 0 400 270]{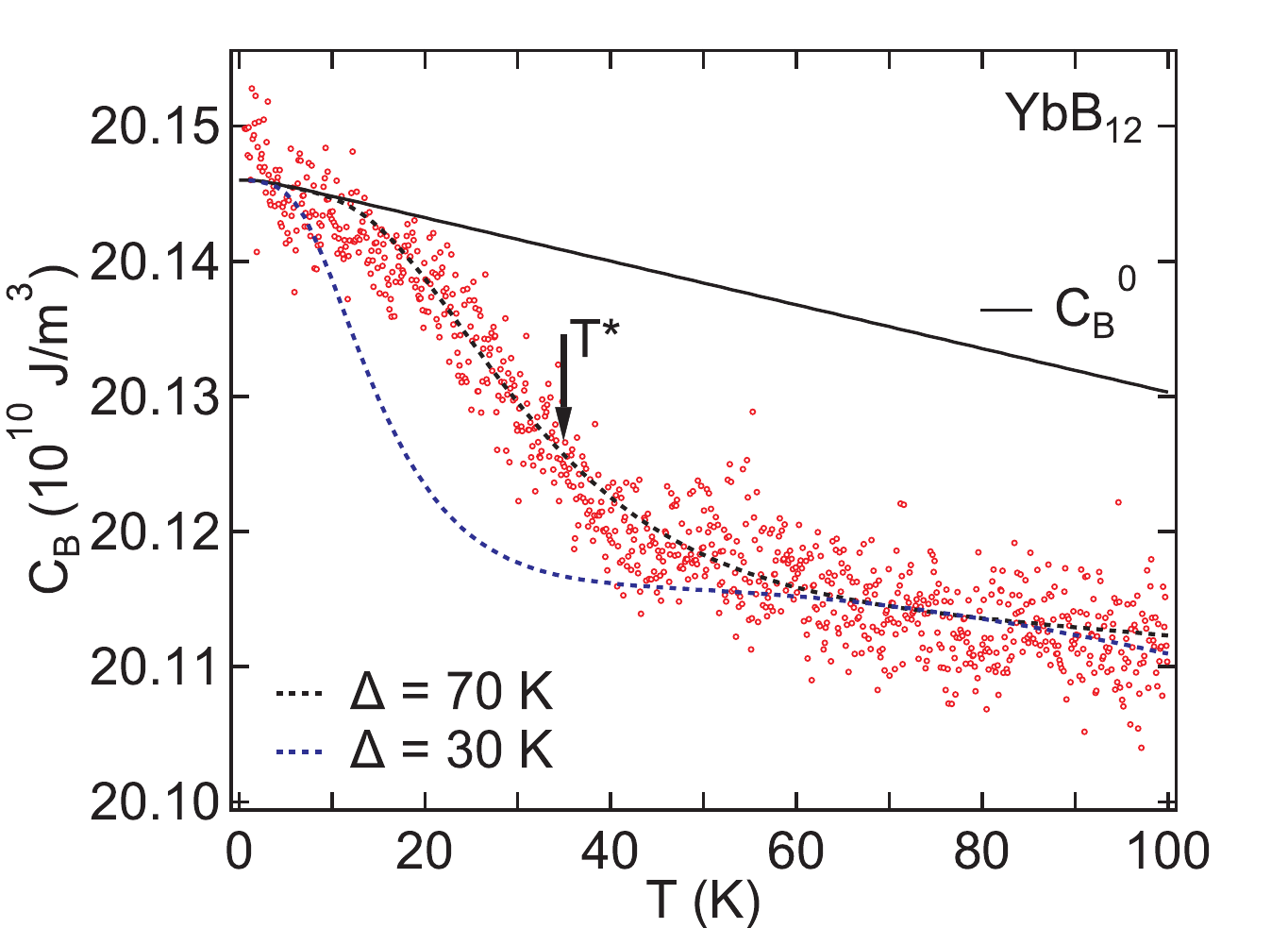}
\end{center}
\caption{
Fit of the bulk modulus $C_\mathrm{B}$ of YbB$_{12}$ by the hexadecapole susceptibility with energy gaps $\Delta = 70$ and $30$ K.
}
\label{Fig11}
\end{figure}
%%%%%%%%%%%%%%%%%%%%%%%%%%%%%%%%%%%%%

Figure \ref{Fig11} shows the fit of bulk modulus $C_\mathrm{B}$ in YbB$_{12}$ by the hexadecapole susceptibility with energy gaps $\Delta = 70$ K and $30$ K.
We cannot describe the curvature change for $\Delta = 30$ K, which corresponds to the activation energy determined by the high-field magnetoresistance
\cite{Sugiyama_JPSJ57}. 
This result indicates that the contribution of the larger gap to the elasticity is dominant in zero field in YbB$_{12}$.

%%%%%%%%%%%%%%%%%%%%%%%%%%%%%%


\begin{thebibliography}{99}

\bibitem{Kasaya_JMMM31} %%% 
M. Kasaya, F. Iga, K. Negishi, S. Nakai, and T. Kasuya,
J. Mag. Mag. Mat. \textbf{31}-\textbf{34}, 437 (1983).
%https://doi.org/10.1016/0304-8853(83)90312-8

\bibitem{Susaki_PRL77}
T. Susaki, A. Sekiyama, K. Kobayashi, T. Mizokawa, A. Fujimori, M. Tsunekawa, T. Muro, T. Matsushita, S. Suga, H. Ishii, T. Hanyu, A. Kimura, H. Namatame, M. Taniguchi, T. Miyahara, F. Iga, M. Kasaya, and H. Harima,
Phys. Rev. Lett. \textbf{77}, 4269 (1996).
%https://doi.org/10.1103/PhysRevLett.77.4269

\bibitem{Nemkovski_PRL99}%%neutron {https://journals.aps.org/prl/abstract/10.1103/PhysRevLett.99.137204}
K. S. Nemkovski, J.-M. Mignot, P. A. Alekseev, A. S. Ivanov, E. V. Nefeodova, A. V. Rybina, L.-P. Regnault, F. Iga, and T. Takabatake, 
Phys. Rev. Lett. \textbf{99}, 137204 (2007).
%https://doi.org/10.1103/PhysRevLett.99.137204

\bibitem{Kanai_JPSJ84}%ground-state%%  {https://journals.jps.jp/doi/10.7566/JPSJ.84.073705}
Y. Kanai,T. Mori, S. Naimen, K. Yamagami, H. Fujiwara, A. Higashiya, T. Kadono, S. Imada, T. Kiss, A. Tanaka, K. Tamasaku, M. Yabashi, T. Ishikawa, F. Iga, and A. Sekiyama, 
J. Phys. Soc. Jpn. \textbf{84}, 073705 (2015).
%https://doi.org/10.7566/JPSJ.84.073705

\bibitem{Ikushima_PhysB281}
K. Ikushima, Y. Kato, M. Takigawa, F. Iga, S. Hiura, and T. Takabatake,
Physica B \textbf{281}-\textbf{282}, 274 (2000).
%https://doi.org/10.1016/S0921-4526(99)00815-7

\bibitem{Kasaya_JMMM47}%%%
M. Kasaya, F. Iga, M. Takigawa, and T. Kasuya,
J. Mag. Mag. Mat. \textbf{47} \& \textbf{48}, 429 (1985).
%https://doi.org/10.1016/0304-8853(85)90458-5

\bibitem{Iga_JMMM177}%%%
F. Iga, N. Shimizu, and T. Takabatake,
J. Mag. Mag. Mat. \textbf{177}-\textbf{181}, 337 (1998).
%https://doi.org/10.1016/S0304-8853(97)00493-9

\bibitem{Yamaguchi_PRB79}%determining valence
J. Yamaguchi, A. Sekiyama, S. Imada, H. Fujiwara, M. Yano, T. Miyamachi, G. Funabashi, M. Obara, A. Higashiya, K. Tamasaku, M. Yabashi, T. Ishikawa, F. Iga, T. Takabatake, and S. Suga,
Phys. Rev. B \textbf{79}, 125121 (2009).
%https://doi.org/10.1103/PhysRevB.79.125121

\bibitem{Saso_JPSJ72}%band calculation
T. Saso and H. Harima,
J. Phys. Soc. Jpn. \textbf{72}, 1131 (2003).
%https://doi.org/10.1143/JPSJ.72.1131

\bibitem{Ohashi_PRB70}%DOS calculation
T. Ohashi, A. Koga, S. I. Suga, and N. Kawakami,
Phys. Rev. B \textbf{70}, 245104 (2004).
%https://doi.org/10.1103/PhysRevB.70.245104

\bibitem{Sugiyama_JPSJ57}
K. Sugiyama, F. Iga, M. Kasaya, T. Kasuya, and M. Date,
J. Phys. Soc. Jpn. \textbf{57}, 3946 (1988).
%https://doi.org/10.1143/JPSJ.57.3946

\bibitem{Iga_JMMM76} %specific heat two band model
F. Iga, M. Kasaya, and T. Kasuya,
J. Mag. Mag. Mat. \textbf{76} \& \textbf{77}, 156 (1988).
%https://doi.org/10.1016/0304-8853(88)90349-6

\bibitem{Takeda_PRB73} %HRPES {https://journals.aps.org/prb/abstract/10.1103/PhysRevB.73.033202}
Y. Takeda, M. Arita, M. Higashiguchi, K. Shimada, H. Namatame, M. Taniguchi, F. Iga, and T. Takabatake,
Phys. Rev. B \textbf{73}, 033202 (2006).
%https://doi.org/10.1103/PhysRevB.73.033202

\bibitem{Iga_JPhysConfSer200}
F. Iga, K. Suga, K. Takeda, S. Michimura, K. Murakami, T. Takabatake, and K. Kindo,
J. Phys.: Conf. Ser. \textbf{200}, 012064 (2010).
%10.1088/1742-6596/200/1/012064

%\bibitem{Matsuda_Crystals}
%Y. H. Matsuda, Y. Kakita, and F. Iga,
%Crystals \textbf{10}, 26 (2020).
%https://doi.org/10.3390/cryst10010026

\bibitem{Terashima_JPSJ86} %high field magnetization {https://journals.jps.jp/doi/10.7566/JPSJ.86.054710}
T. T. Terashima, A. Ikeda, Y. H. Matsuda, A. Kondo, K. Kindo, and F. Iga,
J. Phys. Soc. Jpn. \textbf{86}, 054710 (2017).
%https://doi.org/10.7566/JPSJ.86.054710

\bibitem{Matsuda_JPhys51}%%field dependence of valence
Y. H. Matsuda, Y. Murata, T. Inami, K. Ohwada, H. Nojiri, K. Ohoyama, N. Katoh, Y. Murakami, F. Iga, T. Takabatake, A, Mitsuda, and H. Wada,
J. Phys.:Conf. Ser. \textbf{51}, 111 (2006).
%https://doi.org/10.1088%2F1742-6596%2F51%2F1%2F111

\bibitem{Terashima_PRL120}%%%field induced Kondo metal
T. T. Terashima, Y. H. Matsuda, Y. Kohama, A. Ikeda, A. Kondo, K. Kindo, and F. Iga,
Phys. Rev. Lett. \textbf{120}, 257206 (2018).
%https://doi.org/10.1103/PhysRevLett.120.257206

\bibitem{Xiang_Science362}
Z. Xiang, Y. Kasahara, T. Asaba, B. Lawson, C. Tinsman, L. Chen, K. Sugimoto, S. Kawaguchi, Y. Sato, G. Li, S. Yao, Y. L. Chen, F. Iga, J. Singleton, Y. Matsuda, and L. Li,
Science \textbf{362}, 65 (2018).
%10.1126/science.aap9607 

\bibitem{Luthi Phys. Ac.}
B. L\"uthi,
\textit{Physical Acoustics in the Solid State} (Springer, Berlin, 2005).
%https://doi.org/10.1007/978-3-540-72194-9

\bibitem{Tamaki_JPhysC18}%%%Sm_3_Se_4_
A. Tamaki, T. Goto, S. Kunii, T. Suzuki, T. Fujimura, and T. Kasuya,
J. Phys. C: Solid State Phys. \textbf{18}, 5849 (1985). 
%https://doi.org/10.1088%2F0022-3719%2F18%2F31%2F017

\bibitem{Nemoto_PRB61}%%%Sm_3_Te_4_
Y. Nemoto, T. Goto, A. Ochiai, and T. Suzuki,
Phy. Rev. B \textbf{61}, 12050 (2000).
%https://doi.org/10.1103/PhysRevB.61.12050

\bibitem{Goto_PRB59}%%%Yb_4_As_3_
T. Goto, Y. Nemoto, A. Ochiai, and T. Suzuki,
Phys. Rev. B \textbf{59}, 269 (1999).
%https://doi.org/10.1103/PhysRevB.59.269

\bibitem{Nakamura_JPSJ60_SmB6}%%%SmB_6_
S. Nakamura, T. Goto, M. Kasaya, and S. Kunii,
J. Phys. Soc. Jpn. \textbf{60}, 4311 (1991).
%https://doi.org/10.1143/JPSJ.60.4311

\bibitem{Luthi_JMMM63}%deformation theory and experiments 
B. L\"uthi and M. Yoshizawa,
J. Mag. Mag. Mat. \textbf{63} \& \textbf{64}, 274 (1987).
%https://doi.org/10.1016/0304-8853(87)90586-5

\bibitem{Thalmeier_JPhysC20}
P. Thalmeier, 
J. Phys. C: Solid State Phys. \textbf{20}, 4449 (1987).
%https://doi.org/10.1088%2F0022-3719%2F20%2F28%2F010

\bibitem{Keller_PRB41}%%%Theoretical study, c-f hybridization, elastic softening 
J. Keller, R. Bulla, Th. H\"ohn, and K. W. Becker,
Phys. Rev. B \textbf{41}, 1878 (1990).
%https://doi.org/10.1103/PhysRevB.41.1878

\bibitem{Rout_PhysicaB367}%%%theoretical study of hybridization and electron-phonon interaction
G. C. Rout, M. S. Ojha, and S. N. Behera,
Physica B \textbf{367}, 101 (2005).
%https://doi.org/10.1016/j.physb.2005.06.003

\bibitem{Fujita_JPSJ80}%ultrasonic measurements
T. K. Fujita, M. Yoshizawa, R. Kamiya, H. Mitamura, T. Sakakibara, K. Kindo, F. Iga, I. Ishii, and T. Suzuki,
J. Phys. Soc. Jpn. \textbf{80}, SA084 (2011).
%https://doi.org/10.1143/JPSJS.80SA.SA084

\bibitem{Inui_group}%%%%%%%%%%%‰ž—pŒQ˜_
T. Inui, Y. Tanabe, and Y. Onodera,
\textit{Group Theory and Its Applications in Physics} (Springer, Berlin, 1990).
%10.1007/978-3-642-80021-4

\bibitem{Kuramoto_JPSJ78}
Y. Kuramoto, H. Kusunose, and A. Kiss,
J. Phys. Soc. Jpn. \textbf{78}, 072001 (2009).
%https://doi.org/10.1143/JPSJ.78.072001

\bibitem{Vershni_PRB2}%background elastic constant
Y. P. Varshni,
Phys. Rev. B \textbf{2}, 3952 (1970).
%https://doi.org/10.1103/PhysRevB.2.3952

%\bibitem{InternationalTables}
%\textit{International Tables for Crystallography},
%ed. T. Hahn
%(Kluwer, Dordrecht, 1989) Vol. A, 2nd ed.

\bibitem{Goto_Luthi_review}%review of charge fluctuation and ordering
T. Goto and B. L\"uthi,
Advances in Physics, \textbf{52}, 67 (2003).
%https://doi.org/10.1080/0001873021000057114

%\bibitem{Xiang_Science362}
%Z. Xiang, Y. Kasahara, T. Asaba, B. Lawson, C. Tinsman, L. Chen, K. Sugimoto, S. Kawaguchi, Y. Sato, G. Li, S. Yao, Y. L. Chen, F. Iga, J. Singleton, Y. Matsuda, and L. Li,
%Science \textbf{362}, 65 (2018).

%\bibitem{Sato_NPhys15}
%Y. Sato, Z. Xiang, Y.  Kasahara, T. Taniguchi, S. Kasahara, L. Chen, T. Asaba, C. Tinsman, H. Murayama, O. Tanaka, Y, Mizukami, T. Shibauchi, F. Iga, J. Singleton, L. Li, and Y. Matsuda,
%Nat. Phys. \textbf{15}, 954 (2019).

\bibitem{Kurihara_JPSJ86}%QS in k-space
R. Kurihara, K. Mitsumoto, M. Akatsu, Y. Nemoto, T. Goto, Y. Kobayashi, and S. Sato, 
J. Phys. Soc. Jpn. \textbf{86}, 064706 (2017).
%https://doi.org/10.7566/JPSJ.86.064706

\bibitem{Nakamura_JPSJ60_CeNiSn}%%susceptibility
S. Nakamura, T. Goto, Y. Ishikawa, S. Sakatsume, and M. Kasaya,
J. Phys. Soc. Jpn. \textbf{60}, 2305 (1991).
%https://doi.org/10.1143/JPSJ.60.2305

\bibitem{Luthi_JMMM52}%strain susceptibility
B. L\"uthi,
J. Mag. Mag. Mat. \textbf{52}, 70 (1985).
%https://doi.org/10.1016/0304-8853(85)90228-8

\bibitem{Nakamura_JMMM76}%strain susceptibility of SmB_6
S. Nakamura, T. Goto, T. Fujimura, M. Kasaya, and T. Kasuya,
J. Mag. Mag. Mat. \textbf{76}\&\textbf{77}, 312 (1988).
%https://doi.org/10.1016/0304-8853(88)90407-6

\bibitem{Iga_PhysB186}%pressure dependence of activation energy
F. Iga, M. Kasaya, H. Suzuki, Y. Okayama, H. Takabatake, and N. Mori,
Physica B \textbf{186}, 419 (1993).
%https://doi.org/10.1016/0921-4526(93)90591-S

\bibitem{Mizumaki_JPhysConfSer176}
M. Mizumaki, S. Tsutsui, and F. Iga,
J. Phys.: Conf. Ser. \textbf{176}, 012034 (2009).
%https://doi.org/10.1088%2F1742-6596%2F176%2F1%2F012034

\bibitem{Matsuda_KPS62}
Y. H. Matsuda, T. Nakamura, K. Kuga, and S. Nakatsuji,
J. Korean Phys. Soc. \textbf{62}, 1778 (2013).
%https://doi.org/10.3938/jkps.62.1778

\bibitem{Aoki_PRL71}
H. Aoki, S. Uji, A. K. Albessard, and Y. \=Onuki,
Phys. Rev. Lett. \textbf{71}, 2110 (1993).
%https://doi.org/10.1103/PhysRevLett.71.2110

\bibitem{Matsuda_PRB86}
Y. H. Matsuda, T. Nakamura, J. L. Her, S. Michimura, T. Inami, K. Kindo, and T. Ebihara,
Phys. Rev. B \textbf{86}, 041109(R) (2012).
%https://doi.org/10.1103/PhysRevB.86.041109

\bibitem{Matsuda_SCES2013}
Y. H. Matsuda, J.-L. Her, S. Michimura, T. Inami, T. Ebihara, and H. Amitsuka,
JPS Conf. Proc. \textbf{3}, 011044 (2014).
%https://doi.org/10.7566/JPSCP.3.011044

\bibitem{Moll_NatComm6}
P. J. W. Moll, B. Zeng, L. Balicas, S. Geleski, F F. Balakirev, E. D. Bauer, and F. Ronning,
Nat. Commun. \textbf{6}, 6663 (2015).
%https://doi.org/10.1038/ncomms7663

\bibitem{Ronning_Nature548}
F. Ronning, T. Helm, K. R. Shirer, M. D. Bachmann, L. Balicas, M. K. Chan, B. J. Ramshaw, R. D. McDonald, F. F. Balakirev, M. Jaime, E. D. Bauer, and P. J. W. Moll,
Nature (London) \textbf{548}, 313 (2017).
%10.1038/nature23315 

\bibitem{Rosa_PRL122}
P. F. S. Rosa, S. M. Thomas, F. F. Balakirev, E. D. Bauer, R. M. Fernandes, J. D. Thompson, F. Ronning, and M. Jaime,
Phys. Rev. Lett. \textbf{122}, 016402 (2019).
%https://doi.org/10.1103/PhysRevLett.122.016402

\bibitem{Kurihara_PRB101}
R. Kurihara, A. Miyake, M. Tokunaga, Y. Hirose, and R. Settai,
Phys. Rev. B \textbf{101}, 155125 (2020).
%https://doi.org/10.1103/PhysRevB.101.155125

\bibitem{Iga_AIP8}
F. Iga, K. Yokomichi, W. Matsuhra, H. Nakayama, A. Kondo, K. Kindo, and H. Yoshizawa,
AIP Advances \textbf{8}, 101335 (2018).
%https://doi.org/10.1063/1.5045793 

\bibitem{Watanabe_JPSJ89}
S. Watanabe,
J. Phys. Soc. Jpn. \textbf{89}, 073702 (2020).
%https://doi.org/10.7566/JPSJ.89.073702

\bibitem{Hutchings}%%CEF
M. T. Hutchings,
Solid State Physics \textbf{16}, 227 (1964).
%https://doi.org/10.1016/S0081-1947(08)60517-2


\end{thebibliography}
\end{document}